\documentclass[11pt]{article}
\usepackage{amssymb,amsmath,amsfonts}
\usepackage{graphicx}
\usepackage{lscape}
\usepackage[dvips]{color}
\usepackage[round]{natbib}
\usepackage{float}
\usepackage[colorlinks=true, urlcolor=blue, citecolor=blue, breaklinks]{hyperref}

\newcommand{\blue}{\color[rgb]{0,0,1}}

\numberwithin{equation}{section}
\numberwithin{figure}{section}
\numberwithin{table}{section}

\allowdisplaybreaks

\begin{document}   

\baselineskip 5mm

\thispagestyle{empty}

\begin{center}
{\LARGE {\em Quantile Least Squares\/}: A Flexible Approach 
for Robust}
\\[1ex]
{\LARGE Estimation and Validation of Location-Scale Families}

\vspace{15mm}

{\large\sc
Mohammed Adjieteh\footnote[1]{~Mohammed Adjieteh, Ph.D., ASA, is 
an Assistant Professor in the Department of Mathematical Sciences, 
Appalachian State University, Boone, NC 28608, USA. ~~ {\em e-mail\/}: 
~{\blue\tt adjietehma@appstate.edu}}}  

\vspace{1mm}

{\large\em Appalachian State University}

\vspace{10mm}

{\large\sc  
Vytaras Brazauskas\footnote[2]{
~{\sc Corresponding Author}: Vytaras Brazauskas, Ph.D., ASA,
is a Professor in the Department of Mathematical Sciences,  
University of Wisconsin-Milwaukee, P.O. Box 413, Milwaukee, 
WI 53201, USA. ~~ {\em e-mail\/}: ~{\blue\tt vytaras@uwm.edu}}}  

\vspace{1mm}

{\large\em University of Wisconsin-Milwaukee}

\vspace{15mm}

{\footnotesize To appear in {\em Statistics and Computing}}
\\
{\scriptsize ( {\em Submitted\/}: ~March 11, 2024 ~~
{\em Revised\/}: ~February 14, 2025 ~~
{\em Accepted\/}: ~April 24, 2025 )}
\end{center}

\vspace{-1mm}

\begin{quote}
{\bf\em Abstract\/}.
~In this paper, the problem of robust estimation and validation 
of location-scale families is revisited. The proposed methods 
exploit the joint asymptotic normality of sample quantiles (of 
{\em i.i.d.\/} random variables) to construct the ordinary and 
generalized least squares estimators of location and scale 
parameters. These {\em quantile least squares\/} (QLS) estimators 
are easy to compute because they have explicit expressions, their 
robustness is achieved by excluding extreme quantiles from the 
least-squares estimation, and efficiency is boosted by using 
as many non-extreme quantiles as practically relevant. 
The influence functions of the QLS estimators are specified 
and plotted for several location-scale families. They closely 
resemble the shapes of some well-known influence functions yet 
those shapes emerge automatically (i.e., do not need to be
specified). The joint asymptotic normality of the proposed 
estimators is established and their finite-sample properties 
are explored using simulations. Also, computational costs of 
these estimators, as well as those of MLE, are evaluated for 
sample sizes $n = 10^6, 10^7, 10^8, 10^9$. For model validation, 
two goodness-of-fit tests are constructed and their performance 
is studied using simulations and real data. In particular, for 
the daily stock returns of Google over the years 2020-2023, both 
tests strongly support the logistic distribution assumption and 
reject other bell-shaped competitors.

\vspace{4mm}

{\bf\em Keywords\/}. ~Goodness-of-Fit; Least Squares; Quantiles; 
Relative Efficiency; Robustness.
\end{quote}

\newpage

\baselineskip 7mm

\setcounter{page}{1}

\section{Introduction}

The problem of robust estimation of location-scale families can 
be traced back to the seminal works of 
\citet[][]{t60},
\citet[][]{h64},
and
\citet[][]{h68}.
Since then, numerous robust methods for this problem have been 
proposed in the literature; they are summarized in the books of
\citet[][]{hrrs86},
\citet[][]{mmy06},
and \citet[][]{hr09}.
While at first it might seem like the topic is exhausted and 
fully ``solved'', in this paper we argue that it is worthwhile 
revisiting it. In particular, connections with {\em best linear 
unbiased estimators\/} or BLUE, based on strategically selected 
order statistics, can be exploited and studied from various 
theoretical and practical perspectives: robustness, efficiency, 
model validation (goodness of fit), and computational cost. 
All of this within the same framework.

The literature on BLUE methods for location-scale families, 
which are constructed out of a few order statistics, goes 
back to 
\citet[][]{m46}. 
Since then, numerous papers on parameter estimation, hypothesis 
testing, optimal spacings, simulations, and applications have 
been published. A very short list of contributions to this 
area includes: a first comprehensive review of estimation 
problems by \citet[][]{sg62} (and many specialized papers 
by at least one of these authors);
estimation of parameters of the Cauchy distribution by
\citet[][]{c70} (and multiple related papers by the same 
author and his co-authors) and
\citet[][]{c74};
a relatively recent review of this literature by
\citet[][]{au98} (and numerous technical papers by at least 
one of these authors).
A common theme in many of the papers in this field is to show 
that for various distributions, highly-efficient estimators 
can be constructed using less than ten order statistics. 
Also, when the number of order statistics is fixed, then 
the optimal spacings, according to the asymptotic relative 
efficiency criterion, can be determined. Computational ease 
and robustness of such estimators are often mentioned, but 
to the best of our knowledge no formal studies of robustness 
that specify breakdown points and influence functions have 
been pursued. Interestingly, those optimal (most efficient) 
estimators usually include order statistics that are very 
close to the extreme levels of 0 or 1, making the estimators 
practically nonrobust.

Further, given that location-scale estimation is a special 
case of linear regression, the estimators that are designed 
for robust linear regression with non-normal or contaminated 
errors could be employed. In this line of research, the 
estimators based on a selected type of divergence (e.g., 
$\gamma$-divergence or density power divergence) play
a prominent role and exhibit strong robustness-efficiency 
properties. See, for example, 
\citet[][]{fe08}, \citet[][]{kf17}, and \citet[][]{kf23}. 
However, as \citet[][]{racp20} demonstrate the power divergence 
estimators can be recast as $S$-estimators (with Tukey's biweight), 
which exhibit similar robustness-efficiency trade-offs. Furthermore, 
a different yet quite promising approach in this context is that 
based on the criterion of {\em quantile least squares\/} (QLS), 
which was shown to be effective for robust parameter estimation 
of the $g$-and-$h$ distributional family \citep[][]{xic14}.

In this paper, we use the QLS criterion and propose an original 
estimation method for location-scale families, which relies on 
a linear regression between empirical and theoretical quantiles. 
The novel contribution of the proposed methodology lies in the 
exploitation of the asymptotic structure of the covariance-variance 
matrix to construct more accurate estimators. These estimators 
offer attractive robustness-efficiency trade-offs and exhibit 
other appealing properties, such as their direct applicability 
to any location-scale family (including highly-skewed distributions)
and low computational costs.

The rest of the paper is organized as follows. We first link 
the QLS criterion with BLUE methods for location-scale families 
and thus introduce two types of estimators: ordinary QLS (oQLS) 
and generalized QLS (gQLS). The results in Sections 2 and 3 show 
that the gQLS estimator outperforms the oQLS in terms of the 
robustness-efficiency trade-offs, and hence should be preferred. 
However, the oQLS estimator, being a special case of the gQLS, 
can be useful for finding initial values of the parameters in 
situations when the gQLS has to be computed recursively. Further, 
besides studying small-sample properties of the gQLS estimators 
under clean and contaminated data scenarios (Section 4), we also
evaluate computational costs of gQLS, MLE, and two robust 
$M$-estimators for sample sizes $n = 10^6, 10^7, 10^8, 10^9$. 
In addition, two goodness-of-fit tests are constructed (Section 3.5) 
and their performance is studied using simulations (Section 4.4) 
and real data (Section 5). Finally, a summary of the paper and 
concluding remarks are provided in Section 6.

\section{Quantile Least Squares}

In this section, a general formulation of the least squares estimators
based on sample quantiles is presented, and their asymptotic robustness 
and relative efficiency properties are specified. 

Suppose a sample of {\em independent and identically distributed\/} 
({\em i.i.d.\/}) continuous random variables, $X_1, \ldots, X_n$, 
with the cumulative distribution function (cdf) $F$, probability 
density function (pdf) $f$, and quantile function (qf) $F^{-1}$, 
is observed. Let the cdf, pdf, and qf be given in a parametric form, 
and suppose that they are indexed by an $m$-dimensional parameter 
$\mbox{\boldmath $\theta$} = (\theta_1, \ldots, \theta_m)$.
Further, let $X_{(1)} \leq \cdots \leq X_{(n)}$ denote the ordered 
sample values. The empirical estimator of the $p$th population 
quantile is the corresponding sample quantile
$X_{(\lceil n p \rceil)} = \widehat{F}^{-1}(p)$, where 
$\lceil \cdot \rceil$ denotes the rounding up operation. Also, 
throughout the paper the notation ${\cal AN}$ stands for 
``asymptotically normal.''

\subsection{Regression Estimation}

To specify a regression framework, we first recall the joint
asymptotic normality result of sample quantiles. (The following 
theorem is slightly edited to match the context of the current 
paper.)

\medskip

\noindent
{\bf Theorem 2.1} ~ [ \citet[][p.80, Theorem B]{s02a} ]
\\
{\em Let $0 < p_1 < \cdots < p_k < 1$. Suppose that $F$ has a density 
$f$ in neighborhoods of $F^{-1}(p_1), \ldots, F^{-1}(p_k)$ and that 
$f$ is positive and continuous at $F^{-1}(p_1), \ldots, F^{-1}(p_k)$.
Then the $k$-variate vector of empirical quantiles
$\big( \widehat{F}^{-1}(p_1), \ldots, \widehat{F}^{-1}(p_k) \big)$ 
is asymptotically normal with the mean vector 
$\big( F^{-1}(p_1), \ldots, F^{-1}(p_k) \big)$ 
and $k \times k$ covariance-variance matrix with elements
$\sigma_{ij}/n$, where
\begin{equation}
\sigma_{ij} = \frac{p_i (1-p_j)}{f(F^{-1}(p_i)) f(F^{-1}(p_j))} 
\qquad
\mbox{for ~} i \leq j
\label{emp-sigma}
\end{equation}
and $\sigma_{ij} = \sigma_{ji} \;$ for $i > j$.
}

\medskip

For large sample size and general $F$, this result can be interpreted
as a nonlinear regression model with normally distributed error terms. 
That is,
\begin{equation}
\widehat{F}^{-1}(p_i) ~=~ F^{-1}(p_i) + \varepsilon_i,
\qquad
i = 1, \ldots, k,
\label{nl-reg0}
\end{equation}
where the error term 
$\mbox{\boldmath $\varepsilon$} = 
\left( \varepsilon_1, \ldots, \varepsilon_k \right)$ is 
${\cal AN} \big( \mbox{\bf 0}, \, \mbox{\boldmath $\Sigma$} / n \big)$ 
with the elements of $\mbox{\boldmath $\Sigma$}$ given by \eqref{emp-sigma}. 
Since $F^{-1}(p_i)$ is a function of $\mbox{\boldmath $\theta$} = 
(\theta_1, \ldots, \theta_m)$, the number of quantiles ($k$) should be 
at least as large as the number of parameters ($m$). Then, the least 
squares problem can be formulated as follows:
\begin{equation}
\mbox{minimize} 
\quad
\sum_{i=1}^k \left( \widehat{F}^{-1}(p_i) - F^{-1}(p_i) \right)^2 
\quad
\mbox{with respect to $\theta_1, \ldots, \theta_m$}.
\label{nl-reg1}
\end{equation}
In general, \eqref{nl-reg1} is a challenging computational problem that 
requires numerical optimization algorithms. Moreover, the objective 
function may have many local minima and even the global minimum may 
produce a biased estimate. But as was demonstrated by 
\citet[][Section 2.1]{xic14}
for the $g$-and-$h$ distributional family, this problem can be solved 
with rapidly converging algorithms, and its solution possesses several 
desirable properties: consistency, asymptotic normality, bounded 
influence functions, positive breakdown point. We also notice that 
using similar arguments to those of 
\citet[][]{xic14}
the equivalent theoretical properties can be established for other 
parametric distributions, which will be discussed in Sections 2.2-2.3. 
Further, it will be shown in Section 3 that for location-scale families 
and their variants, the nonlinear regression model \eqref{nl-reg0} 
becomes a linear regression model with (approximately) normally 
distributed error terms whose covariance-variance matrix has a 
convenient structure. As a result, the latter problem has explicit 
solutions with known theoretical properties.

\subsection{Robustness Properties}

The quantile least squares (QLS) estimator found by solving 
\eqref{nl-reg1} can be viewed as an {\em indirect\/} estimator, 
robust inferential properties of which are provided by 
\citet[][]{gdl00} and \citet[][]{gr03}. 
Using those general results and the arguments of \citet[][]{xic14}
several properties of the QLS estimator can be stated. First, as is 
clear from the choice of quantile confidence levels,
\[
0 < a = p_1 < p_2 < \cdots < p_{k-1} < p_k = b < 1,
\]
the order statistics with the index less than $\lceil n a \rceil$ 
and more than $\lceil n b \rceil$ play no role in estimation of the 
regression model \eqref{nl-reg0}. This implies that the QLS estimator
is globally robust with the (asymptotic) breakdown point equal to:
\begin{equation}
\mbox{BP} ~=~ 
\min \left\{ \mbox{LBP}, \mbox{UBP} \right\} 
~=~ \min \left\{ a, 1-b \right\} > 0.
\label{nl-bp}
\end{equation}
Note that when the underlying probability distribution $F$ is not 
symmetric it makes sense to consider {\em lower\/} (LBP) and 
{\em upper\/} (UBP) breakdown points separately. For more 
details on the relevance of LBP and UBP in applications, 
see 
\citet[][]{bs00} and \citet[][]{s02b}.
Second, the influence function (IF) of the QLS estimator for
$\mbox{\boldmath $\theta$}$ is directly related to the influence 
functions of ``data'', i.e., the selected sample quantiles 
$\widehat{F}^{-1}(p_1), \ldots, \widehat{F}^{-1}(p_k)$:
\[
\mbox{IF} \big( x, \widehat{F}^{-1}(p_i) \big) ~=~
\frac{p_i - \mbox{\bf\large 1} \{ x \leq F^{-1}(p_i) \}}{f(F^{-1}(p_i))},
\qquad i = 1, \ldots, k,
\]
where $-\infty < x < \infty$ and $\mbox{\bf\large 1} \{ \cdot \}$ 
denotes the indicator function. Specifically, the IF of
$\widehat{\mbox{\boldmath $\theta$}}$ is given by
\begin{equation}
\mbox{IF} \big( x, \widehat{\mbox{\boldmath $\theta$}} \big) ~=~
(\mathbf{X' X})^{-1} \mathbf{X'} 
\left( \mbox{IF} \big( x, \widehat{F}^{-1}(p_1) \big), \ldots,
\mbox{IF} \big( x, \widehat{F}^{-1}(p_k) \big) \right)',
\label{nl-if}
\end{equation}
where $\mathbf{X} = \big[ X_{ij} \big]_{k \times m} = 
\left[ \frac{\partial F^{-1}(p_i)}{\partial \theta_j} 
\right]_{k \times m}$, and is bounded because $p_1 = a > 0$
and $p_k = b < 1$.

\subsection{Asymptotic Relative Efficiency}

To start with, the model assumptions used in Theorem 1 of 
\citet[][Section 2.1]{xic14} can be broadened to include other 
parametric models. Then, repeating the arguments used by these authors 
to prove the theorem, it can be established that in general the QLS 
estimator is consistent and ${\cal AN}$ with the mean vector 
$\mbox{\boldmath $\theta$}$ and $m \times m$ 
covariance-variance matrix 
\begin{equation}
\dfrac{1}{n} \, (\mathbf{X' X})^{-1} \mathbf{X'} 
\mbox{\boldmath $\Sigma$}
\mathbf{X} (\mathbf{X' X})^{-1} ,
\label{nl-an}
\end{equation}
where $\mathbf{X}$ is defined as in \eqref{nl-if} and the elements 
of $\mbox{\boldmath $\Sigma$}$ are given by \eqref{emp-sigma}.

Further, under suitable regularity conditions 
\citep[][Section 4.2.2]{s02a},
the maximum likelihood estimator (MLE) is ${\cal AN}$ with the mean 
vector $\mbox{\boldmath $\theta$}$ and $m \times m$ covariance-variance 
matrix $\frac{1}{n} \, \mathbf{I}^{-1}$,
where $\mathbf{I}$ is the Fisher information matrix. Since MLE is
the most efficient ${\cal AN}$ estimator (i.e., its asymptotic 
variance attains the Cram{\'{e}}r-Rao bound), its performance 
can serve as a benchmark for the QLS estimator. In particular, 
the following {\em asymptotic relative efficiency\/} (ARE) 
criterion will be used:
\begin{equation}
\mbox{ARE} \, \big( \mbox{QLS}, \, \mbox{MLE} \big) ~=~ 
\left( \frac{\mbox{det} \left[ \mathbf I^{-1} \right]}
{\mbox{det} 
\big[
(\mathbf{X' X})^{-1} \mathbf{X'} 
\mbox{\boldmath $\Sigma$}
\mathbf{X} (\mathbf{X' X})^{-1}
\big]} 
\right)^{1/m} ,
\label{are}
\end{equation}
where `det' stands for the determinant of a square matrix
\citep[][Section 4.1]{s02a}.

\section{Location-Scale Families}

In this section, the proposed methodology is worked out for 
location-scale families. Several such families are listed in 
Section 3.1. Two QLS-type estimators are developed in Section 3.2. 
Further, efficiency and robustness properties of the new estimators 
are established in Sections 3.3 and 3.4, respectively. Finally, 
in Section 3.5, we analyze model residuals and explore its 
goodness-of-fit properties. 

\subsection{Preliminaries}

The pdf $f$, cdf $F$, and the qf $F^{-1}$ of the location-scale 
family are given by:
\begin{equation}
f(x) = \frac{1}{\sigma} f_* \left( \frac{x-\mu}{\sigma} \right), 
\qquad
F(x) = F_* \left( \frac{x-\mu}{\sigma} \right),
\qquad
F^{-1}(u) = \mu + \sigma F_*^{-1}(u),
\label{func}
\end{equation}
where $-\infty < x < \infty$ (or depending on the distribution, $x$ can 
be restricted to some interval), $0 < u < 1$, $-\infty < \mu < \infty$ 
is the location parameter, and $\sigma > 0$ is the scale parameter.
The functions $f_*$, $F_*$, $F_*^{-1}$ represent pdf, cdf, qf,
respectively, of the standard location-scale family (i.e., $\mu=0$, 
$\sigma=1$). Choosing $\mu$ or $\sigma$ known, the location-scale
family is reduced to either the {\em scale\/} or {\em location\/} 
family, respectively. 

In Table 3.1, we list key facts for several location-scale families. 
The selected distributions include typical symmetric bell-shaped 
densities, with domains on all real numbers (e.g., Cauchy, Laplace, 
Logistic, Normal, Student's $t$), and few asymmetric densities with 
varying domains (e.g., Exponential, Gumbel, L{\'{e}}vy). In the latter 
group, the Gumbel pdf is defined on all real numbers but is slightly 
skewed; this distribution plays an important role in extreme value 
theory. Two-parameter Exponential and L{\'{e}}vy densities are highly 
skewed and have domains $(\mu, \, \infty)$. They represent examples 
when the aforementioned regularity conditions are not satisfied, 
due to the presence of $\mu$. 
Both distributions are widely used in applications and have many 
probabilistic connections. For example, the L{\'{e}}vy distribution 
is directly related to the following well-known distributions: 
Inverse Gamma, Stable, Folded Normal.

\begin{center}
{\sc Table 3.1.} Key probabilistic formulas and information 
for selected location-scale families.

\medskip

\begin{tabular}{|c|c|c|c|}
\hline
Probability & Standard {\sc pdf} & Standard {\sc qf} & 
Information Matrix \\[-0.5ex]
Distribution & $f_*(z)$ & $F_*^{-1}(u)$ & 
$\mathbf{I_*} \, ( = \sigma^2 \times \mathbf{I} )$ \\
\hline
\hline
Cauchy & $\dfrac{1}{\pi (1 + z^2)}$ & 
$\tan (\pi (u-0.5))$ & 
$\begin{bmatrix}
\frac{1}{2} & 0 \\
0 & \frac{1}{2} \\
\end{bmatrix}$ \\
Laplace & $0.5 \, e^{-|z|}$ & 
$\left\{ 
\begin{array}{cl} 
 \ln (2 u), & u \leq 0.5
\\[0.25ex]
 -\ln (2 (1-u)), & u > 0.5
\end{array}
\right.
$
& 
$\begin{bmatrix}
1 & 0 \\
0 & 1 \\
\end{bmatrix}$ \\
Logistic & $\dfrac{e^{-z}}{(1+e^{-z})^2}$ & 
$-\ln (1/u-1)$ & 
$\begin{bmatrix}
\frac{1}{3} & 0 \\
0 & \frac{3+\pi^2}{9} \\
\end{bmatrix}$ \\
Normal & $\frac{1}{\sqrt{2 \pi}} \, e^{-z^2/2}$ & 
$\Phi^{-1}(u)$ & 
$\begin{bmatrix}
1 & 0 \\
0 & 2 \\
\end{bmatrix}$ \\
Student's $t$ &  
$\frac{1}{\sqrt{\nu}} \frac{\Gamma \left( \frac{\nu+1}{2} \right)}
{\Gamma \left( \frac{1}{2} \right) \Gamma \left( \frac{\nu}{2} \right)}
\left( 1+\frac{z^2}{\nu} \right)^{-\frac{\nu+1}{2}}$ & 
$F^{-1}_{t_{\nu}}(u)$ & 
$\begin{bmatrix}
\frac{\nu+1}{\nu+3} & 0 \\
0 & \frac{2 \nu}{\nu+3} \\
\end{bmatrix}$ \\
\hline
Exponential & $e^{-z}$, {\footnotesize $z > 0$} & $- \ln (1-u)$ & 
$1$ {\footnotesize (for $\sigma$; $\mu$ is known)} \\
Gumbel & $\displaystyle \exp \{-z - e^{-z} \}$ & 
$- \ln (- \ln(u) )$ & 
$\begin{bmatrix}
1 & \gamma-1 \\
\gamma-1 & \frac{\pi^2}{6} + (\gamma-1)^2 \\
\end{bmatrix}$ \\
L{\'{e}}vy  & 
$\frac{1}{\sqrt{2 \pi}} \, z^{-3/2} \, e^{-(2z)^{-1}}$, {\footnotesize $z > 0$} & 
$\big( \Phi^{-1}(1-u/2) \big)^{-2}$ &
$\frac{1}{2}$ {\footnotesize (for $\sigma$; $\mu$ is known)} \\[1ex]
\hline
\multicolumn{4}{l}{} \\[-3ex]
\multicolumn{4}{l}{\footnotesize {\sc Note 1}: 
~$F^{-1}_{t_{\nu}}$ is the qf of the standard $t$ distribution 
with $\nu$ degrees of freedom ($\nu > 0$ is known).} \\[-0.5ex]
\multicolumn{4}{l}{\footnotesize {\sc Note 2}: ~$\gamma = -\Gamma'(1) 
\approx 0.5772$ is the Euler-Mascheroni constant.} \\
\end{tabular}
\end{center}

It is worthwhile mentioning that there exist numerous variants of the 
location-scale family such as folded distributions (e.g., Folded Normal, 
Folded Cauchy) or log-location-scale families (e.g., Lognormal, Pareto 
type $I$). Since their treatment requires only suitable parameter-free 
transformation of the data variable, the estimators developed in this 
paper will work for those distributions as well.

\subsection{Parameter Estimation}

Incorporating expressions \eqref{func} into the model \eqref{nl-reg0} 
yields a linear regression model:
\begin{equation}
\mathbf{Y} ~=~ \mathbf{X} \mbox{\boldmath $\beta$} + 
\mbox{\boldmath $\varepsilon$},
\label{l-reg}
\end{equation}
where
$\mathbf{Y} = \left( \widehat{F}^{-1}(p_1), \ldots, \widehat{F}^{-1}(p_k) \right)'$,
~{\boldmath $\beta$} $= (\mu, \sigma)'$, and
~$\mbox{\boldmath $\varepsilon$} = 
\left( \varepsilon_1, \ldots, \varepsilon_k \right)'$ is 
${\cal AN} \big( \mbox{\bf 0}, \, \sigma^2 \mbox{\boldmath $\Sigma_*$} / n \big)$.
The entries of {\boldmath $\Sigma_*$} are defined by \eqref{emp-sigma}, 
but now they are completely {\em known\/} because $f$ and $F^{-1}$ are replaced 
with $f_*$ and $F_*^{-1}$, respectively. The design matrix $\mathbf{X}$ is 
defined as in \eqref{nl-if} and has simple entries:
\begin{equation}
\mathbf{X} ~=~ 
\begin{bmatrix}
\frac{\partial F^{-1}(p_1)}{\partial \mu} & \cdot & \cdot & \cdot & 
\frac{\partial F^{-1}(p_k)}{\partial \mu} \\[1ex]
\frac{\partial F^{-1}(p_1)}{\partial \sigma} & \cdot & \cdot & \cdot & 
\frac{\partial F^{-1}(p_k)}{\partial \sigma} \\
\end{bmatrix}' ~=~
\begin{bmatrix}
1 & \cdot & \cdot & \cdot & 1 \\[1ex]
F_*^{-1}(p_1) & \cdot & \cdot & \cdot & F_*^{-1}(p_k) \\
\end{bmatrix}'.
\label{matrixX}
\end{equation}

Solving \eqref{nl-reg1} for the model \eqref{l-reg} leads to 
the {\em ordinary\/} least squares estimator
\begin{equation}
\widehat{\mbox{\boldmath $\beta$}}_{\mbox{\tiny oQLS}} ~=~
(\mathbf{X' X})^{-1} \mathbf{X'} \mathbf{Y}
\label{oQLS}
\end{equation}
which is ${\cal AN}$ with the mean vector {\boldmath $\beta$} 
$= (\mu, \sigma)'$ and $2 \times 2$ covariance-variance matrix
\begin{equation}
\dfrac{\sigma^2}{n} \, (\mathbf{X' X})^{-1} \mathbf{X'} 
\mbox{\boldmath $\Sigma_*$}
\mathbf{X} (\mathbf{X' X})^{-1} ,
\label{l-an-o}
\end{equation}
where $\mathbf{X}$ is given by \eqref{matrixX}.

Further, the oQLS solution \eqref{oQLS} implicitly assumes that 
{\boldmath $\Sigma_*$} is the $k \times k$ identity matrix, which 
might be a sensible assumption for the non-linear regression model 
\eqref{nl-reg1} because the resulting estimator is consistent while 
the computational complexity is significantly reduced. In general, 
however, such a simplification decreases the estimator's efficiency. 
Since for the linear regression model \eqref{l-reg}, 
{\boldmath $\Sigma_*$} is known, ARE of oQLS can be improved by 
employing the {\em generalized\/} least squares estimator
\begin{equation}
\widehat{\mbox{\boldmath $\beta$}}_{\mbox{\tiny gQLS}} ~=~
(\mathbf{X'} \mbox{\boldmath $\Sigma_*^{-1}$} \mathbf{X})^{-1} 
\mathbf{X'} \mbox{\boldmath $\Sigma_*^{-1}$} \mathbf{Y}
\label{gQLS}
\end{equation}
which is ${\cal AN}$ with the mean vector {\boldmath $\beta$} 
$= (\mu, \sigma)'$ and $2 \times 2$ covariance-variance matrix
\begin{equation}
\dfrac{\sigma^2}{n} \, 
( \mathbf{X'} \mbox{\boldmath $\Sigma_*^{-1}$} \mathbf{X} )^{-1} .
\label{l-an-g}
\end{equation}

Finally, note that if a one-parameter family -- location or scale -- 
needs to be estimated, the formulas \eqref{oQLS}--\eqref{l-an-g} 
still remain valid, but the design matrix \eqref{matrixX} would
be a column of 1's (for location) or a column of $F_*^{-1}(p)$'s 
(for scale).

\subsection{Relative Efficiency Studies}

To see how much efficiency is sacrificed when one uses 
$\widehat{\mbox{\boldmath $\beta$}}_{\mbox{\tiny oQLS}}$
instead of
$\widehat{\mbox{\boldmath $\beta$}}_{\mbox{\tiny gQLS}}$, 
we will compute \eqref{are} for several distributions of 
Table 3.1. In view of \eqref{l-an-o} and \eqref{l-an-g}, 
the ARE formula \eqref{are} is now given by
\[
\mbox{ARE} \, \big( \mbox{oQLS}, \, \mbox{MLE} \big) ~=~ 
\left( \frac{\mbox{det} \left[ \mathbf I_*^{-1} \right]}
{\mbox{det} 
\big[
(\mathbf{X' X})^{-1} \mathbf{X'} 
\mbox{\boldmath $\Sigma_*$}
\mathbf{X} (\mathbf{X' X})^{-1}
\big]} 
\right)^{1/2}
\]
and
\[
\mbox{ARE} \, \big( \mbox{gQLS}, \, \mbox{MLE} \big) ~=~ 
\left( \frac{\mbox{det} \left[ \mathbf I_*^{-1} \right]}
{\mbox{det} 
\big[
( \mathbf{X'} \mbox{\boldmath $\Sigma^{-1}_*$} \mathbf{X} )^{-1}
\big]} 
\right)^{1/2} ,
\]
where $\mathbf I_*$ is specified in Table 3.1. For one-parameter 
families (location or scale), the covariance-variance matrices in 
the ARE formulas get reduced to scalars and the exponents become 1.

These ARE expressions are functions of $k$, the number of selected 
sample quantiles, therefore the choice of $p_1, \ldots, p_k$ (with
$k \geq 2$) is important. As mentioned earlier, our top priority is 
to identify QLS estimators that offer the best trade-off between 
robustness and efficiency. Thus, to keep the breakdown points 
positive and influence functions bounded, we first fix $a = p_1 > 0$ 
and $b = p_k < 1$ and then make the remaining $p_i$'s equally spaced:
\begin{equation}
p_i = a + \frac{i-1}{k-1} (b-a), 
\qquad
i = 1, \ldots, k.
\label{quants}
\end{equation}

It is clear that choosing larger $a$ and smaller $b$ yields higher 
robustness, while choosing larger $k$ improves efficiency. But there 
are practical limits to the efficiency improvement. As can be seen 
from Figure 3.1, the (pointwise) ARE curves become almost flat for 
$k > 10$ making the efficiency gains negligible. This holds true 
irrespectively of the underlying location-scale family. 
On the other hand, choosing gQLS over oQLS gives a major boost to 
ARE, especially for the heavier-tailed distributions such as Gumbel, 
Laplace, and Cauchy. Also, the seesaw pattern of the ARE curve (for 
location and to a lesser degree for location-scale) for the Laplace 
distribution can be explained as follows: for $a=1-b$ and $k$ odd, 
one of the $p_i$'s is always equal to 0.50 resulting in the 
selection of the sample median, which in this case is MLE for 
$\mu$ and thus full efficiency is attained. 

\bigskip

\noindent
{\bf Note 3.1} ~If the number of selected quantiles $k$ is fixed, 
then a non-uniform spacing of $p_i$'s could potentially improve 
the efficiency of the QLS estimators while delivering the same 
levels of robustness. However, the approach \eqref{quants} resolves 
this problem by simply increasing $k$. The only limitation is the 
number of observations that are left between the levels $a$ and $b$. 
(For example, for $n=100$, $a=0.10$, and $b=0.90$, there would be 
80 observations left to include in QLS estimation.) The main advantage 
of the uniform selection of $p_i$'s over other options is its simplicity 
and universal applicability. \hfill $\Box$

\bigskip

Further, to demonstrate that increasing $k$ yields no substantial 
gains in efficiency, in Table 3.2 we list AREs of the generalized 
QLS estimators for Cauchy, Gumbel, Laplace, Logistic, and Normal 
distributions, when $k = 15, \, 20, \, 25$. As is evident from 
the table, choosing $k=25$ over $k=15$ results in $\sim 1\%$
improvement of AREs. Similar magnitude improvements can be 
observed when the extreme quantile levels $a$ and $b$ are 
changed from $(0.02, 0.98)$ to $(0.05, 0.95)$ to $(0.10, 0.90)$.
In view of this discussion and keeping in mind that $k$ should 
be odd (see the ARE entries for Laplace), we can say that the 
choice of $k=15$ would suffice in most situations. However, 
in order to not squander much efficiency at some unusual 
location-scale distributions, $k=25$ is a safer choice.

\begin{center}
\resizebox{165mm}{165mm}{\includegraphics{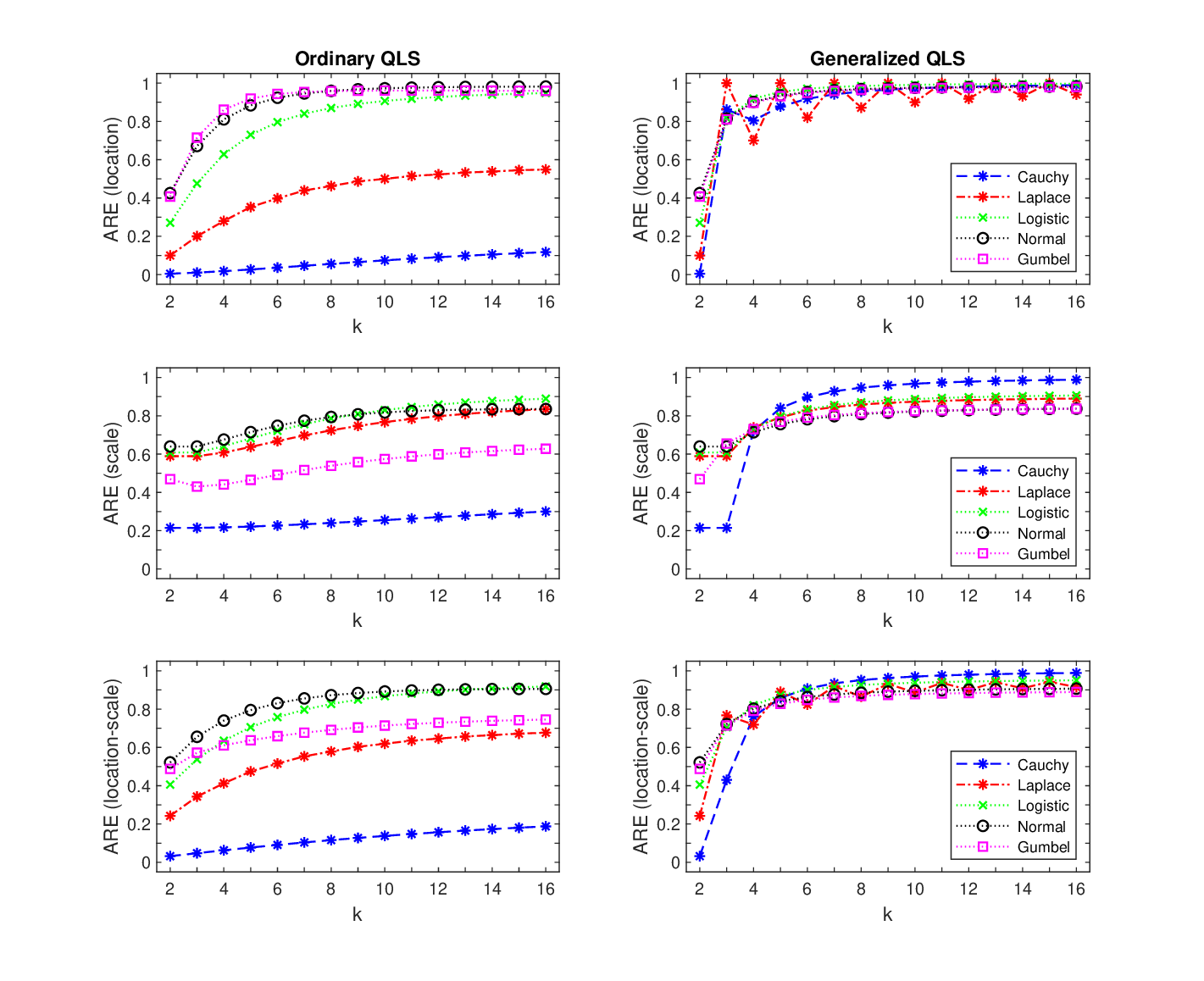}}
\\[-3ex]
{\sc Figure 3.1.} ~AREs of the ordinary and generalized QLS 
estimators of location, scale, and joint 
\\[-1ex]
location-scale parameters for Cauchy, Gumbel, Laplace, Logistic, 
and Normal distributions.
\\[-1ex]
The quantiles are selected according to \eqref{quants} with
$(a, b) = (0.05, 0.95)$ and $k = 2:16$.
\end{center}

Finally, it is tempting to try the brute force approach for 
the oQLS estimators with the hope that it would substantially 
improve ARE. We computed AREs for oQLS when $k$ ranges from 
15 to 200, but found that the improvements are minor. (Those 
computations will not be presented here.) In particular, 
depending on the distribution and how low the ARE value is 
at $k=15$, some tiny improvements are still possible even 
for $k=200$ (Cauchy) but they are leveling off (Logistic). 
More interestingly, for the Laplace, Normal, and Gumbel 
distributions, their AREs reach the peak at some lower 
$k$ and then start slowly declining. This behavior is not 
unexpected because the oQLS estimator, while consistent, 
is based on an incorrect simplifying assumption that 
{\boldmath $\Sigma_*$} is the $k \times k$ identity matrix.
This assumption, combined with the brute force approach, will 
eventually penalize the estimator's performance.

\begin{center}
{\sc Table 3.2.} AREs of the generalized QLS estimators of location, 
scale, and joint
\\[-1ex]
location-scale parameters for Cauchy, Gumbel, Laplace, Logistic, and 
Normal distributions.
\\[-1ex]
The quantiles are selected according to \eqref{quants} with various
$(a, b)$ and $k = 15, \, 20, \, 25$.

\medskip

\begin{tabular}{|c|ccc|ccc|ccc|}
\hline
Probability & 
\multicolumn{3}{|c|}{Location} &
\multicolumn{3}{|c|}{Scale} &
\multicolumn{3}{|c|}{Location-Scale} \\[-0.5ex]
Distribution & $k=15$ & $k=20$ & $k=25$ & $k=15$ & $k=20$ & $k=25$ & 
$k=15$ & $k=20$ & $k=25$ \\
\hline
\multicolumn{10}{l}{$(a, \, b) = (0.02, \, 0.98)$} \\
\hline
Cauchy   & 0.986 & 0.992 & 0.995 & 0.985 & 0.992 & 0.995 & 0.985 & 0.992 & 0.995 \\
Laplace  &   1   & 0.950 &   1   & 0.930 & 0.943 & 0.949 & 0.965 & 0.946 & 0.974 \\
Logistic & 0.996 & 0.998 & 0.998 & 0.938 & 0.951 & 0.958 & 0.966 & 0.974 & 0.978 \\
Normal   & 0.987 & 0.991 & 0.992 & 0.901 & 0.915 & 0.922 & 0.943 & 0.952 & 0.957 \\
Gumbel   & 0.985 & 0.990 & 0.991 & 0.902 & 0.913 & 0.918 & 0.933 & 0.941 & 0.946 \\
\hline
\multicolumn{10}{l}{$(a, \, b) = (0.05, \, 0.95)$} \\
\hline
Cauchy   & 0.988 & 0.993 & 0.995 & 0.987 & 0.993 & 0.995 & 0.987 & 0.993 & 0.995 \\
Laplace  &   1   & 0.953 &   1   & 0.888 & 0.894 & 0.896 & 0.943 & 0.923 & 0.947 \\
Logistic & 0.996 & 0.998 & 0.999 & 0.904 & 0.910 & 0.913 & 0.949 & 0.953 & 0.955 \\
Normal   & 0.982 & 0.984 & 0.985 & 0.836 & 0.841 & 0.843 & 0.906 & 0.909 & 0.911 \\
Gumbel   & 0.979 & 0.981 & 0.982 & 0.836 & 0.840 & 0.842 & 0.888 & 0.892 & 0.893 \\
\hline
\multicolumn{10}{l}{$(a, \, b) = (0.10, \, 0.90)$} \\
\hline
Cauchy   & 0.981 & 0.985 & 0.986 & 0.989 & 0.993 & 0.995 & 0.985 & 0.989 & 0.991 \\
Laplace  &   1   & 0.958 &   1   & 0.796 & 0.798 & 0.799 & 0.892 & 0.874 & 0.894 \\
Logistic & 0.995 & 0.997 & 0.997 & 0.814 & 0.816 & 0.817 & 0.900 & 0.902 & 0.903 \\
Normal   & 0.964 & 0.965 & 0.965 & 0.708 & 0.710 & 0.711 & 0.826 & 0.828 & 0.828 \\
Gumbel   & 0.956 & 0.957 & 0.957 & 0.719 & 0.720 & 0.721 & 0.803 & 0.805 & 0.805 \\
\hline
\end{tabular}
\end{center}

\subsection{Robustness Investigations}

To see what kind of shapes the influence functions of
$\widehat{\mbox{\boldmath $\beta$}}_{\mbox{\tiny oQLS}}$
and
$\widehat{\mbox{\boldmath $\beta$}}_{\mbox{\tiny gQLS}}$
exhibit, we evaluate and plot \eqref{nl-if} for the symmetric 
(Figure 3.2) and asymmetric (Figure 3.3) location-scale families
of Table 3.1. In view of \eqref{oQLS}, \eqref{gQLS}, and 
\eqref{func}, the expression \eqref{nl-if} is now given by
\[
\mbox{IF} 
\big( x, \widehat{\mbox{\boldmath $\beta$}}_{\mbox{\tiny oQLS}} \big) ~=~
(\mathbf{X' X})^{-1} \mathbf{X'} 
\left( \mbox{IF} \big( x, \widehat{F}^{-1}(p_1) \big), \ldots,
\mbox{IF} \big( x, \widehat{F}^{-1}(p_k) \big) \right)'
\]
and
\[
\mbox{IF} 
\big( x, \widehat{\mbox{\boldmath $\beta$}}_{\mbox{\tiny gQLS}} \big) ~=~
(\mathbf{X'} \mbox{\boldmath $\Sigma_*^{-1}$} \mathbf{X})^{-1} 
\mathbf{X'} \mbox{\boldmath $\Sigma_*^{-1}$}  
\left( \mbox{IF} \big( x, \widehat{F}^{-1}(p_1) \big), \ldots,
\mbox{IF} \big( x, \widehat{F}^{-1}(p_k) \big) \right)',
\]
where
$
\mbox{IF} \big( x, \widehat{F}^{-1}(p_i) \big) = \sigma \,
\frac{p_i - \mbox{\bf 1} \{ x \, \leq \, \mu + 
\sigma F_*^{-1}(p_i) \}}{f_*(F_*^{-1}(p_i))},
~ i = 1, \ldots, k.
$
In Figures 3.2 and 3.3, $\mu = 0$ and $\sigma = 1$.

\medskip

\begin{center}
\resizebox{170mm}{130mm}{\includegraphics{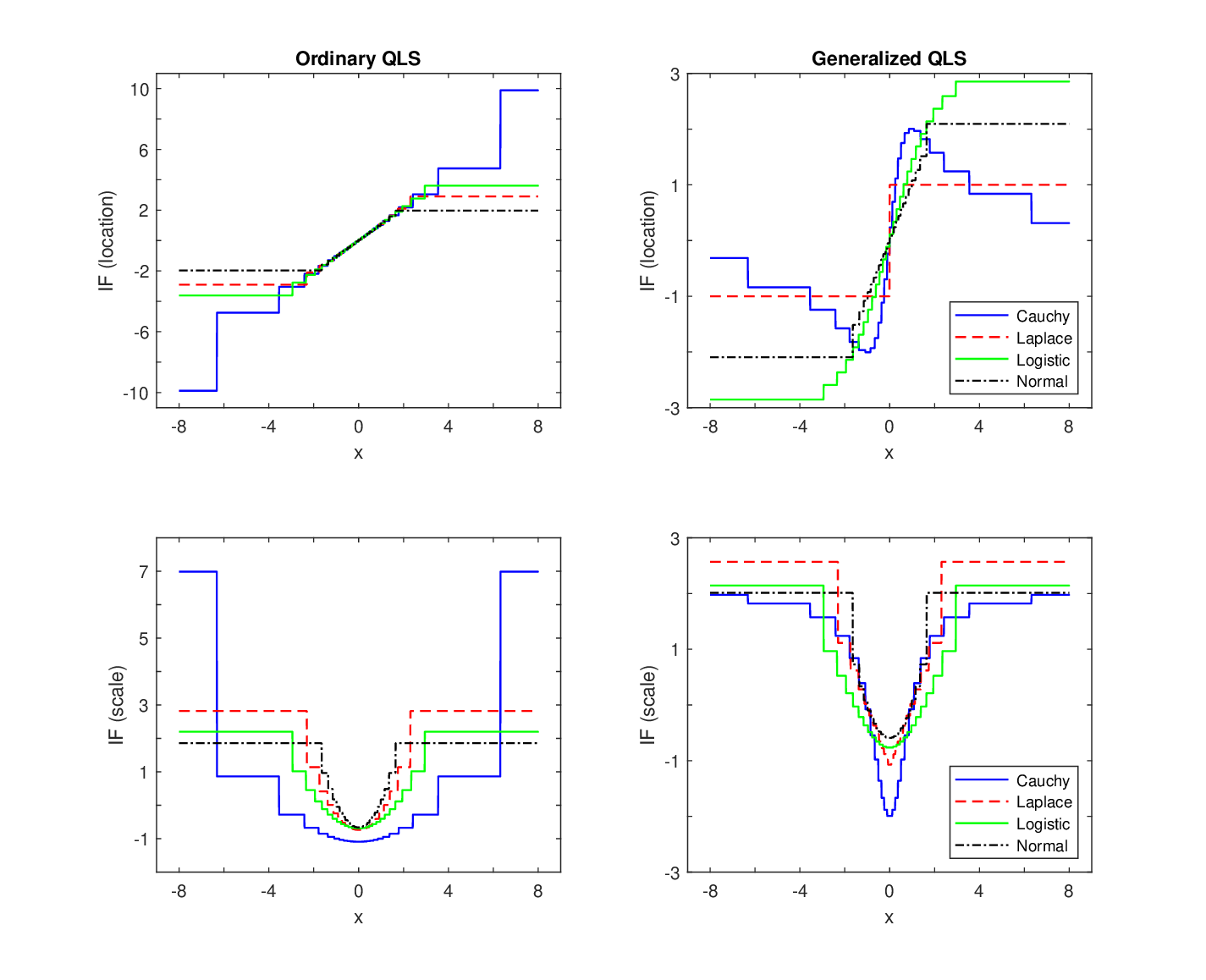}}
\\[-2ex]
{\sc Figure 3.2.} ~Influence functions of the ordinary and 
generalized QLS estimators of 
\\[-1ex]
location and scale parameters for Cauchy, Laplace, Logistic, 
and Normal distributions.
\\[-1ex]
The quantiles are selected according to \eqref{quants} with
$(a, b) = (0.05, 0.95)$ and $k = 25$.
\end{center}

\medskip

In Figure 3.2, we see that the IF shapes of the ordinary QLS 
estimators look familiar. For estimation of $\mu$, the estimators 
act like a stepwise approximation of a trimmed/winsorized mean or 
Huber estimator \citep[][Figure 1, p.105]{hrrs86}. For estimation 
of $\sigma$, they behave like an approximate version of an 
$M$-estimator for scale \citep[][Figure 2, p.123]{hrrs86}.
On the other hand, the generalized QLS estimators demonstrate 
a remarkable flexibility. For estimation of $\sigma$, gQLS shrinks 
the height of the Cauchy IF and keeps the other curves similar to 
those of oQLS. But most impressively, it automatically changes 
the shape of the IF when estimating $\mu$: for Normal and Logistic 
distributions, it acts like a trimmed/winsorized mean; for Laplace, 
it behaves like a median; and for Cauchy, its shape resembles that 
of a Tukey's biweight \citep[][Figure 3, p.151]{hrrs86}.

\medskip

\begin{center}
\resizebox{170mm}{130mm}{\includegraphics{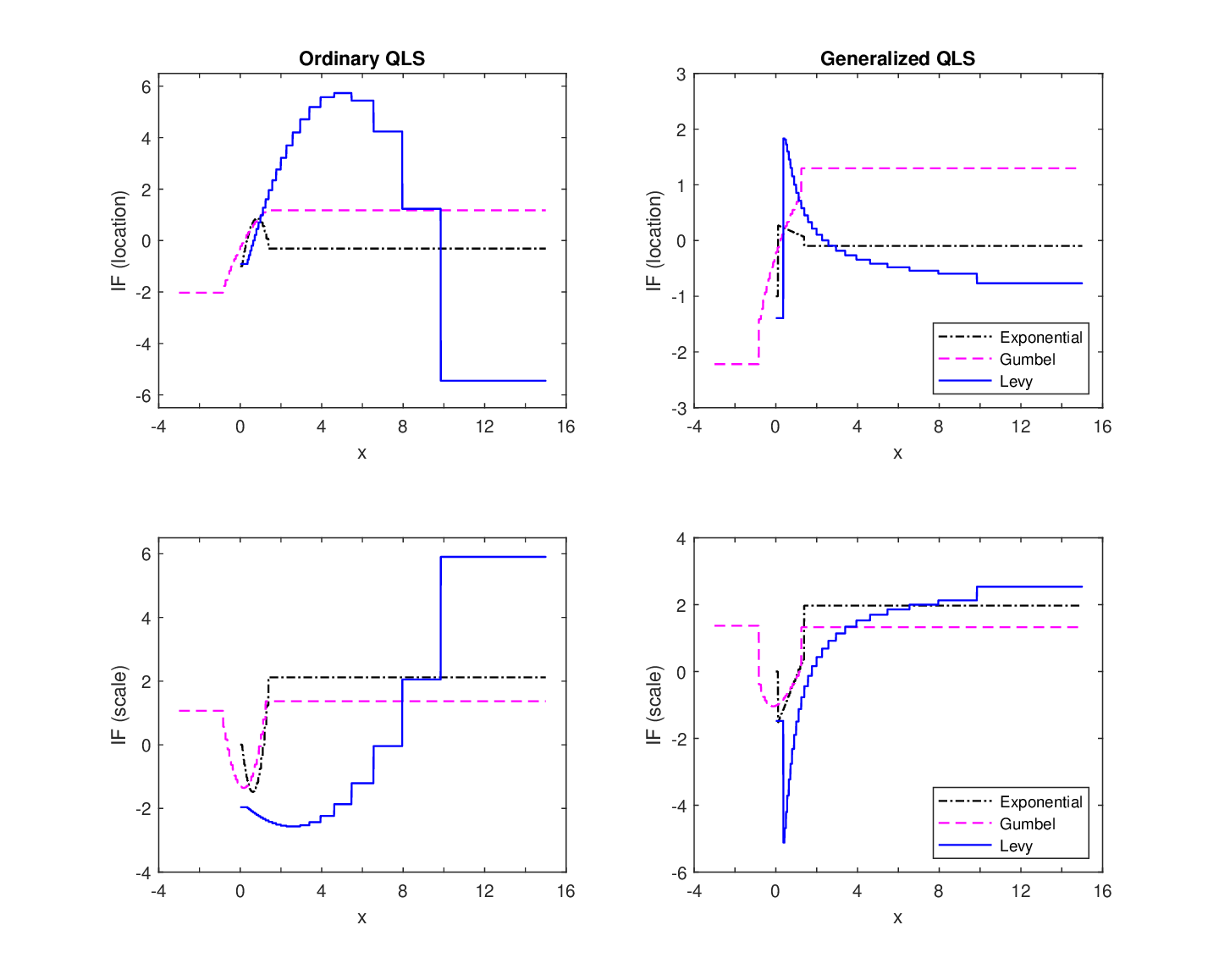}}
\\[-2ex]
{\sc Figure 3.3.} ~Influence functions of the ordinary and 
generalized QLS estimators
\\[-1ex]
of location and scale parameters for Exponential, Gumbel, 
and L{\'{e}}vy distributions.
\\[-1ex]
The quantiles are selected according to \eqref{quants} with
$(a, b) = (0.10, 0.75)$ and $k = 25$.
\end{center}

\medskip

In Figure 3.3, the shapes of IF are predictably non-symmetric. 
For oQLS and gQLS at Gumbel, they are fairly similar to the IFs 
of Normal or Logistic distributions. (Note that by choosing 
$a = 0.10 \ne 0.25 = 1-b$ we made the IF look more symmetric 
than it actually is.)
For Exponential and L{\'{e}}vy, parameter $\mu$ is not the 
``center'' of the pdf anymore; it is the left boundary of 
its support. This fact has an effect on the shapes of IFs. 
For gQLS of $\mu$ and $\sigma$, we see that points near the 
boundary exhibit most dramatic swings of influence. Overall, 
these IFs can be seen as a half of a symmetric family IF.

\bigskip

\noindent
{\bf Note 3.2} ~Based on the analysis presented in this section,
it is clear that the gQLS estimators outperform their oQLS 
counterparts in terms of ARE (except at Normal and perhaps 
Logistic distributions) while offering exactly the same levels 
of LBP and UBP. In terms of computational costs, the two types 
of estimators are essentially equivalent. Therefore, oQLS will 
not be included in further investigations in this paper. 
Nonetheless, this estimator, being a special case of the gQLS, 
can be useful for finding initial values of the parameters in 
situations when the gQLS has to be computed recursively. Think, 
for example, of the gamma distribution when the shape parameter 
$\alpha$ is unknown. 
\hfill $\Box$

\subsection{Model Validation}

For model validation, we consider two goodness-of-fit tests,
both are constructed using 
$\widehat{\mbox{\boldmath $\beta$}}_{\mbox{\tiny gQLS}}$.
The first test is a typical $\chi^2$ test that is based on 
a quadratic form in model residuals. This approach will be 
called ``in-sample validation'' (Section 3.5.1). The second 
test is conceptually similar but is based on a combination 
of the model residuals and additional sample quantiles. 
The inclusion of quantiles that had not been used for 
parameter estimation allows us to make a fair comparison 
among the estimators with different $a$ and $b$. This 
approach will be called ``out-of-sample validation'' 
(Section 3.5.2). 

\subsubsection{In-Sample Validation}

After the parameter estimation step is completed, the predicted 
value of $\mathbf{Y}$ is defined as $\widehat{\mathbf{Y}} ~=~
\mathbf{X} \widehat{\mbox{\boldmath $\beta$}}_{\mbox{\tiny gQLS}}$.
Then the corresponding residuals are
\[
\widehat{\mbox{\boldmath $\varepsilon$}} ~=~ 
\mathbf{Y} - \widehat{\mathbf{Y}} ~=~ \mathbf{Y} - 
\mathbf{X} \widehat{\mbox{\boldmath $\beta$}}_{\mbox{\tiny gQLS}} ~=~
\big( \mbox{\boldmath $\mbox{\bf I}_k$} - \mathbf{X}
(\mathbf{X'} \mbox{\boldmath $\Sigma_*^{-1}$} \mathbf{X})^{-1} 
\mathbf{X'} \mbox{\boldmath $\Sigma_*^{-1}$} 
\big)
\mathbf{Y},
\]
where $\mbox{\boldmath $\mbox{\bf I}_k$}$ is the $k \times k$ 
identity matrix.
Using \eqref{gQLS}, \eqref{l-an-g} and standard statistical 
inference techniques for linear models 
\citep[][Section 12.3]{hmc05}
the following properties can be verified:
\begin{itemize}
  \item $\mathbf{Y}$ has a ${\cal AN}$ 
$\left(
\mathbf{X} \mbox{\boldmath $\beta$}, \;
\dfrac{\sigma^2}{n} \, \mbox{\boldmath $\Sigma_*$} 
\right)$ 
distribution.

  \item $\widehat{\mathbf{Y}}$ has a ${\cal AN}$ 
$\left(
\mathbf{X} \mbox{\boldmath $\beta$}, \;
\dfrac{\sigma^2}{n} \, \mathbf{X}
(\mathbf{X'} \mbox{\boldmath $\Sigma_*^{-1}$} \mathbf{X})^{-1} 
\mathbf{X'} 
\right)$
distribution.

  \item $\widehat{\mbox{\boldmath $\varepsilon$}}$ has a
${\cal AN}$ 
$\left(
\mathbf{0}, \;
\dfrac{\sigma^2}{n} \, \Big( \mbox{\boldmath $\Sigma_*$} - 
\mathbf{X}
(\mathbf{X'} \mbox{\boldmath $\Sigma_*^{-1}$} \mathbf{X})^{-1} 
\mathbf{X'} \Big)
\right)$
distribution.

  \item $\widehat{\mathbf{Y}}$ and 
$\widehat{\mbox{\boldmath $\varepsilon$}}$ are (asymptotically) 
independent.
\end{itemize}
Next, these properties can be exploited to construct a diagnostic
plot (e.g., predicted values {\em versus\/} residuals) and to show
that the quadratic form
\[
Q ~=~ \frac{n}{\sigma^2} 
\left( \mathbf{Y} - \mathbf{X} \mbox{\boldmath $\beta$} \right)'
\mbox{\boldmath $\Sigma_*^{-1}$}
\left( \mathbf{Y} - \mathbf{X} \mbox{\boldmath $\beta$} \right)
\]
has the following orthogonal decomposition:
\[
Q ~=~ Q_1 + Q_2 ~=~ \frac{n}{\sigma^2} 
\left( \mathbf{Y} - \mathbf{X} 
\widehat{\mbox{\boldmath $\beta$}}_{\mbox{\tiny gQLS}} \right)'
\mbox{\boldmath $\Sigma_*^{-1}$}
\left( \mathbf{Y} - \mathbf{X} 
\widehat{\mbox{\boldmath $\beta$}}_{\mbox{\tiny gQLS}} \right) +
\frac{n}{\sigma^2} 
\left( \widehat{\mbox{\boldmath $\beta$}}_{\mbox{\tiny gQLS}} - 
\mbox{\boldmath $\beta$} \right)'
\mathbf{X'} \mbox{\boldmath $\Sigma_*^{-1}$} \mathbf{X}
\left( \widehat{\mbox{\boldmath $\beta$}}_{\mbox{\tiny gQLS}} - 
\mbox{\boldmath $\beta$} \right).
\]
Therefore, since asymptotically $Q$ has a $\chi^2_k$ distribution 
and $Q_2$ has a $\chi^2_2$ distribution, the above decomposition 
implies that $Q_1$ has an approximate $\chi^2_{k-2}$ distribution.

Now, to test the hypotheses 
\[
\left\{ 
\begin{array}{cl}
H_0: & X_1, \ldots, X_n 
\mbox{~ were generated by a location-scale family } F \\
H_A: & X_1, \ldots, X_n 
\mbox{~ were {\em not\/} generated by } F, \\
\end{array}
\right.
\]
the quadratic form $Q_1$ can be utilized as follows. Recall that
$\widehat{\mbox{\boldmath $\beta$}}_{\mbox{\tiny gQLS}} =
(\widehat{\mu}_{\mbox{\tiny gQLS}}, 
\widehat{\sigma}_{\mbox{\tiny gQLS}})'$ is a consistent estimator
of {\boldmath $\beta$}, thus $\widehat{\sigma}^2_{\mbox{\tiny gQLS}}$
converges in probability to $\sigma^2$. Define a test statistic 
\begin{equation}
W ~=~ \frac{n}{\widehat{\sigma}^2_{\mbox{\tiny gQLS}}} 
\left( \mathbf{Y} - \mathbf{X} 
\widehat{\mbox{\boldmath $\beta$}}_{\mbox{\tiny gQLS}} \right)'
\mbox{\boldmath $\Sigma_*^{-1}$}
\left( \mathbf{Y} - \mathbf{X} 
\widehat{\mbox{\boldmath $\beta$}}_{\mbox{\tiny gQLS}} \right).
\label{gof1}
\end{equation}
Since 
$W = \frac{\sigma^2}{\widehat{\sigma}^2_{\mbox{\tiny gQLS}}} \, Q_1$
and 
$\frac{\sigma^2}{\widehat{\sigma}^2_{\mbox{\tiny gQLS}}} \rightarrow 1$ 
(in probability), it follows from Slutsky's Theorem that under $H_0$ 
the test statistic $W$ has an approximate $\chi^2_{k-2}$ distribution. 
Note that a similar goodness-of-fit test was proposed by \citet[][]{au89}, 
but there $\sigma^2$ was estimated by the sample variance, which requires 
that $F$ has a finite variance. The test based on \eqref{gof1} has wider 
applicability (e.g., it works for heavy-tailed distributions such as 
Cauchy) and inherits the robustness properties of 
$\widehat{\mbox{\boldmath $\beta$}}_{\mbox{\tiny gQLS}}$.

\subsubsection{Out-of-Sample Validation}

To compare the goodness of fit of location-scale distributions for 
which $\widehat{\mbox{\boldmath $\beta$}}_{\mbox{\tiny gQLS}}$ are 
computed using different $a$ and $b$ (i.e., $a_1, b_1$ versus 
$a_2, b_2$), we first fix a universal set of sample quantiles. 
That is, select $\mathbf{Y}_{\mbox{\tiny out}} = 
\left( \widehat{F}(p_1^{\mbox{\tiny out}}), \ldots, 
\widehat{F}(p_r^{\mbox{\tiny out}}) \right)'$, 
where 
$p_1^{\mbox{\tiny out}}, \ldots, p_r^{\mbox{\tiny out}}$ can be all 
different from, partially overlapping with, or completely match 
$p_1, \ldots, p_k$ (which are used for parameter estimation). 
Of course, the latter choice simplifies the out-sample-validation 
test to the test of Section 3.5.1. After this selection is made, we 
proceed by mimicking the structure of \eqref{gof1}. The predicted 
value of $\mathbf{Y}_{\mbox{\tiny out}}$ is 
$\mathbf{X_{\mbox{\tiny out}}} 
\widehat{\mbox{\boldmath $\beta$}}_{\mbox{\tiny gQLS}}$
with
\[
\mathbf{X}_{\mbox{\tiny out}} ~=~ 
\begin{bmatrix}
1 & \cdot & \cdot & \cdot & 1 \\[1ex]
F_*^{-1}(p_1^{\mbox{\tiny out}}) & \cdot & \cdot & \cdot & 
F_*^{-1}(p_r^{\mbox{\tiny out}}) \\
\end{bmatrix}',
\]
but $\widehat{\mbox{\boldmath $\beta$}}_{\mbox{\tiny gQLS}} =
(\widehat{\mu}_{\mbox{\tiny gQLS}},\widehat{\sigma}_{\mbox{\tiny gQLS}})'$
is based on 
$\mathbf{Y} = \left( \widehat{F}(p_1), \ldots, \widehat{F}(p_k) \right)'$.
Then the test statistic is
\begin{equation}
W_{\mbox{\tiny out}} ~=~ \frac{n}{\widehat{\sigma}^2_{\mbox{\tiny gQLS}}} 
\left( \mathbf{Y}_{\mbox{\tiny out}} - \mathbf{X}_{\mbox{\tiny out}} 
\widehat{\mbox{\boldmath $\beta$}}_{\mbox{\tiny gQLS}} \right)'
\mbox{\boldmath $\Sigma_{\mbox{\tiny out}}^{-1}$}
\left( \mathbf{Y}_{\mbox{\tiny out}} - \mathbf{X}_{\mbox{\tiny out}} 
\widehat{\mbox{\boldmath $\beta$}}_{\mbox{\tiny gQLS}} \right),
\label{gof2}
\end{equation}
where the elements of {\boldmath $\Sigma_{\mbox{\tiny out}}$} 
are $\sigma_{ij}^{\mbox{\tiny out}} = 
\frac{p_i^{\mbox{\tiny out}} (1-p_j^{\mbox{\tiny out}})}
{f_*(F_*^{-1}(p_i^{\mbox{\tiny out}})) 
f_*(F_*^{-1}(p_j^{\mbox{\tiny out}}))}$ for $~i \leq j~$ 
with $~i, j = 1, \ldots, r$.

Now, unless $p_1^{\mbox{\tiny out}}, \ldots, p_r^{\mbox{\tiny out}}$ 
perfectly match $p_1, \ldots, p_k$ (this case was solved in Section 3.5.1), 
the theoretical derivation of the distribution of $W_{\mbox{\tiny out}}$ 
is a major challenge. Therefore, to estimate the $p$-value associated 
with this test statistic, the following bootstrap procedure can be 
employed.

\begin{figure}[ht!]
\begin{minipage}{17cm}
\begin{enumerate}
  \item[] \underline{\hspace{\linewidth}}

\vspace{-1ex}

  \item[] {\bf ~ Bootstrap Procedure} (for finding 
the $p$-value of \eqref{gof2})

\vspace{-3ex}

  \item[] \underline{\hspace{\linewidth}}

\vspace{-1ex}

  \item[] {\bf Step 1.} Given the original sample, 
$X_1, \ldots, X_n$, the estimates of 
{\boldmath $\beta$} and $W_{\mbox{\tiny out}}$ 
are obtained. Denote them 
$\widehat{\mbox{\boldmath $\beta$}}^o_{\mbox{\tiny gQLS}} =
(\widehat{\mu}^o_{\mbox{\tiny gQLS}},
\widehat{\sigma}^o_{\mbox{\tiny gQLS}})'$
and $\widehat{W}^o_{\mbox{\tiny out}}$.
Remember that 
$\widehat{\mbox{\boldmath $\beta$}}^o_{\mbox{\tiny gQLS}}$
is computed using the quantile levels $p_1, \ldots, p_k$, 
while $\widehat{W}^o_{\mbox{\tiny out}}$ is based on
$p_1^{\mbox{\tiny out}}, \ldots, p_r^{\mbox{\tiny out}}$ and
$\widehat{\mbox{\boldmath $\beta$}}^o_{\mbox{\tiny gQLS}}$.

  \item[] {\bf Step 2.} Generate an {\em i.i.d\/}. sample 
  $X_1^{(b)}, \ldots, X_n^{(b)}$
from $F$ (assumed under $H_0$) using the parameter values
$\widehat{\mbox{\boldmath $\beta$}}^o_{\mbox{\tiny gQLS}} =
(\widehat{\mu}^o_{\mbox{\tiny gQLS}},
\widehat{\sigma}^o_{\mbox{\tiny gQLS}})'$.
Based on this sample, compute
$\widehat{\mbox{\boldmath $\beta$}}^{(b)}_{\mbox{\tiny gQLS}}$
(using $p_1, \ldots, p_k$) and
$\widehat{W}^{(b)}_{\mbox{\tiny out}}$ (using 
$p_1^{\mbox{\tiny out}}, \ldots, p_r^{\mbox{\tiny out}}$ and
$\widehat{\mbox{\boldmath $\beta$}}^{(b)}_{\mbox{\tiny gQLS}}$).

  \item[] {\bf Step 3.} Repeat {\sc Step 2} a $B$ number of times 
(e.g., $B=1000$) and save 
$\widehat{W}^{(1)}_{\mbox{\tiny out}}, \ldots, 
\widehat{W}^{(B)}_{\mbox{\tiny out}}$.

  \item[] {\bf Step 4.} Estimate the $p$-value of \eqref{gof2} by
\[
\widehat{p}_{\mbox{\tiny val}} ~=~ 
\frac{1}{B} \sum_{b=1}^B {\mbox{\large\bf 1}} 
\left\{ \widehat{W}^{(b)}_{\mbox{\tiny out}} > 
\widehat{W}^o_{\mbox{\tiny out}} \right\}
\]
and reject $H_0$ when 
$\widehat{p}_{\mbox{\tiny val}} \leq \alpha$
(e.g., $\alpha = 0.05$).

\vspace{-3ex}

  \item[] \underline{\hspace{\linewidth}}
\end{enumerate}
\end{minipage}
\end{figure}

\section{Simulations}

In this section, we conduct a simulation study with the two-fold 
objective: ($i$) to verify and augment the theoretical properties 
established in Section 3, and ($ii$) to see how the proposed 
estimator, gQLS, compares to two well-established $M$-estimators, 
Huber (H) and Tukey's biweight (T).
We start by describing the study design (Section 4.1). Then we 
explore how the MLE, T, H, and gQLS estimators perform as the sample 
size $n$ increases (Section 4.2), and when data are contaminated with 
outliers (Section 4.3). We finish the study by investigating the power 
of the goodness-of-fit tests against several alternatives (Section 4.4).

\subsection{Study Design}

The study design is based on the following choices.
\begin{figure}[ht!]
\begin{minipage}{17cm}
\begin{enumerate}
  \item[] \underline{\hspace{\linewidth}}

\vspace{-1ex}

  \item[] {\bf ~ Simulation Design}

\vspace{-3ex}

  \item[] \underline{\hspace{\linewidth}}

\vspace{-1ex}

\begin{itemize}
  \item {\em Location-scale families\/} ($F_0$ with $\mu=0$ and $\sigma=1$).
~Cauchy, Exponential, Gumbel, Laplace, L{\'{e}}vy, Logistic, Normal.

  \item {\em Estimators\/}. ~MLE, T, H, and gQLS (abbreviated `g'). 
For the T and H estimators, the tuning constants are the default 
values in $R$: $c=4.685, \, k=1.345$ (for T) and $c=k=1.345$ (for H).
For the `g' estimators, the quantiles are selected according 
to \eqref{quants} with
$(a_1, b_1) = (0.02, 0.98)$,
$(a_2, b_2) = (0.05, 0.95)$,
$(a_3, b_3) = (0.10, 0.90)$
and $k = 25$ (in all cases).

  \item {\em Contaminating distribution\/} ($G$ in the contamination model
$F_{\varepsilon} = (1-\varepsilon) \, F_0 + \varepsilon \, G$, where $F_0$ 
is a location-scale family). ~$\mbox{Normal} \, (\mu^* = 1, \sigma^* = 3)$.

  \item {\em Levels of contamination\/}.
~$\varepsilon = 0, \, 0.03, \, 0.05, \, 0.08$.
 
  \item {\em Goodness-of-fit tests\/} (at $\alpha = 0.05$). 
~Based on $W$, given by \eqref{gof1}, and 
$W_{\mbox{\tiny out}}$, given by \eqref{gof2}.

  \item {\em Quantile levels for model validation\/}. 
~$p_1^{\mbox{\tiny out}} = 0.01, \, p_2^{\mbox{\tiny out}} = 0.03, \ldots, 
\, p_{49}^{\mbox{\tiny out}} = 0.97, \, p_{50}^{\mbox{\tiny out}} = 0.99$.
  
  \item {\em Sample sizes\/}. ~$n = 10^2, 10^3$ (and 
$n = 10^6, 10^7, 10^8, 10^9$ for computational time evaluations).
  
  \item {\em Number of bootstrap samples\/}. ~$B = 10^3$.

  \item {\em Number of Monte Carlo runs\/}. ~$M = 10^4$.  
\end{itemize}

\vspace{-4ex}

  \item[] \underline{\hspace{\linewidth}}

\end{enumerate}
\end{minipage}
\end{figure}

For any given distribution, we generate $M = 10^4$ random samples 
of a specified length $n$. For each sample, we estimate $\mu$ and 
$\sigma$ of the distribution using the MLE, T, H, and gQLS estimators. 
The results are then presented using boxplots and few summarizing 
statistics.

\subsection{From Small to Big Data}

The boxplots of the estimators under consideration are presented 
in Figures 4.1 (for Normal and Cauchy) and 4.2 (for Exponential 
and L{\'{e}}vy). Barring a few exceptions, most estimators are 
correctly calibrated, i.e., they are centered at $\mu = 0$ for 
location and $\sigma = 1$ for scale, and shrink toward the 
respective targets according to the rate of $n^{1/2}$. 
The latter statement can be illustrated by reporting 
the values of the ratio
$\sqrt{\mbox{\sc mse}}$ (at $n=100$)$\big/
\sqrt{\mbox{\sc mse}}$ (at $n=1000$) for each estimator, 
and few selected distributions.
\begin{itemize}
  \item For Normal, the ratios are: 
3.17 ($\widehat{\mu}_{\mbox{\tiny MLE}}$),
3.17 ($\widehat{\mu}_{\mbox{\tiny T}}$),
3.17 ($\widehat{\mu}_{\mbox{\tiny H}}$),
3.24 ($\widehat{\mu}_{\mbox{\tiny g1}}$),
3.19 ($\widehat{\mu}_{\mbox{\tiny g2}}$),
3.18 ($\widehat{\mu}_{\mbox{\tiny g3}}$);
and
3.20 ($\widehat{\sigma}_{\mbox{\tiny MLE}}$),
3.12 ($\widehat{\sigma}_{\mbox{\tiny T}}$),
3.12 ($\widehat{\sigma}_{\mbox{\tiny H}}$),
3.17 ($\widehat{\sigma}_{\mbox{\tiny g1}}$),
3.18 ($\widehat{\sigma}_{\mbox{\tiny g2}}$),
3.16 ($\widehat{\sigma}_{\mbox{\tiny g3}}$).

  \item For Cauchy, the ratios are: 
3.21 ($\widehat{\mu}_{\mbox{\tiny MLE}}$),
3.18 ($\widehat{\mu}_{\mbox{\tiny T}}$),
2.72 ($\widehat{\mu}_{\mbox{\tiny H}}$),
3.32 ($\widehat{\mu}_{\mbox{\tiny g1}}$),
3.25 ($\widehat{\mu}_{\mbox{\tiny g2}}$),
3.21 ($\widehat{\mu}_{\mbox{\tiny g3}}$);
and
3.16 ($\widehat{\sigma}_{\mbox{\tiny MLE}}$),
1.00 ($\widehat{\sigma}_{\mbox{\tiny T}}$),
0.07 ($\widehat{\sigma}_{\mbox{\tiny H}}$),
3.30 ($\widehat{\sigma}_{\mbox{\tiny g1}}$),
3.33 ($\widehat{\sigma}_{\mbox{\tiny g2}}$),
3.30 ($\widehat{\sigma}_{\mbox{\tiny g3}}$).
\end{itemize}
Note that according to the asymptotic results of Section 3.2, 
these ratios are expected to fall around 
$\sqrt{1000/100} \approx 3.16$. 
The incorrect behavior of T and H under Cauchy is due to their 
being tuned for the normal distribution. (This should be 
interpreted as a warning against ``automatically'' using 
off-the-shelf software products.)
Similar conclusions can be drawn for distributions in Figure 4.2.

\begin{center}
\resizebox{160mm}{100mm}{\includegraphics{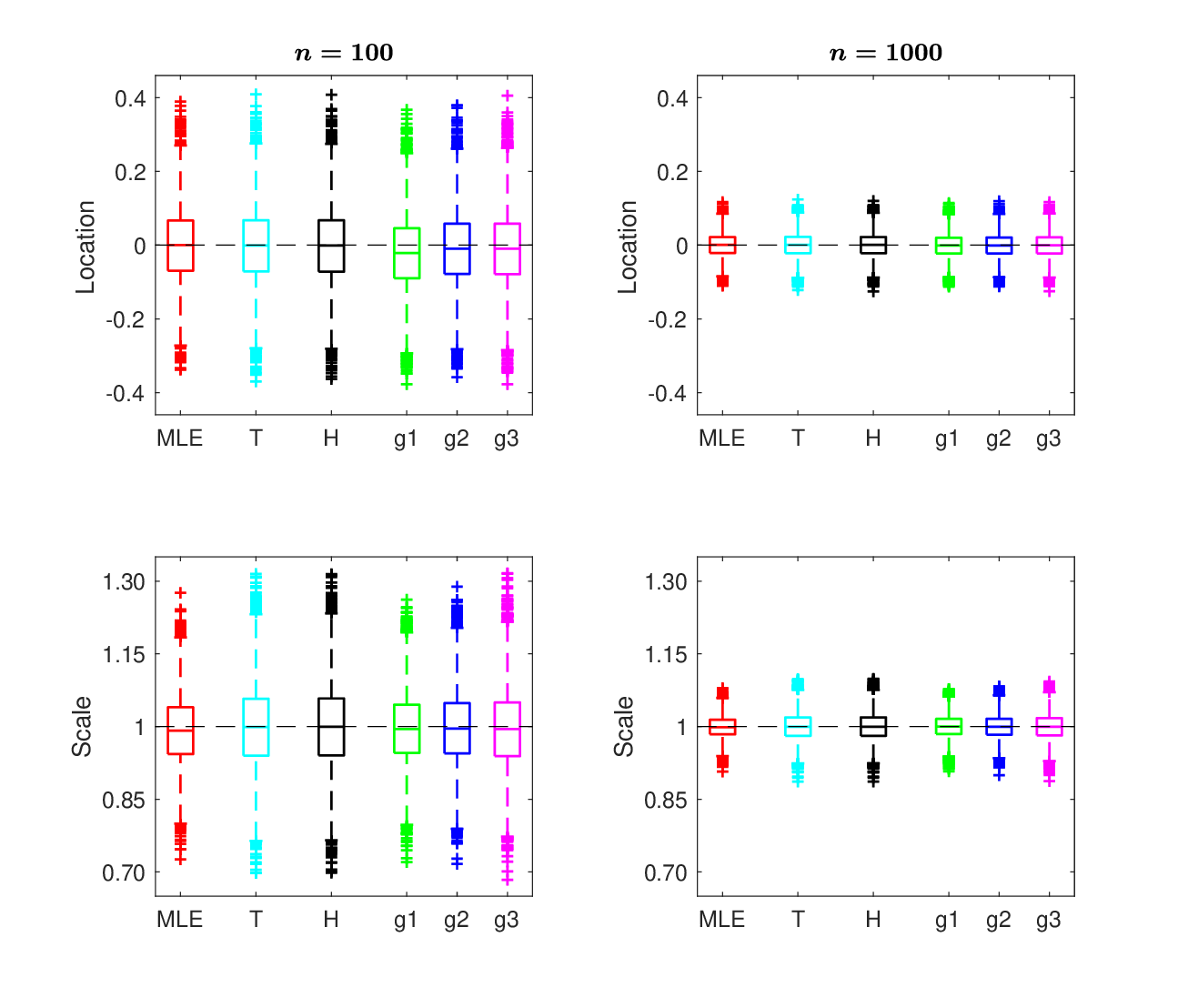}}
\resizebox{160mm}{100mm}{\includegraphics{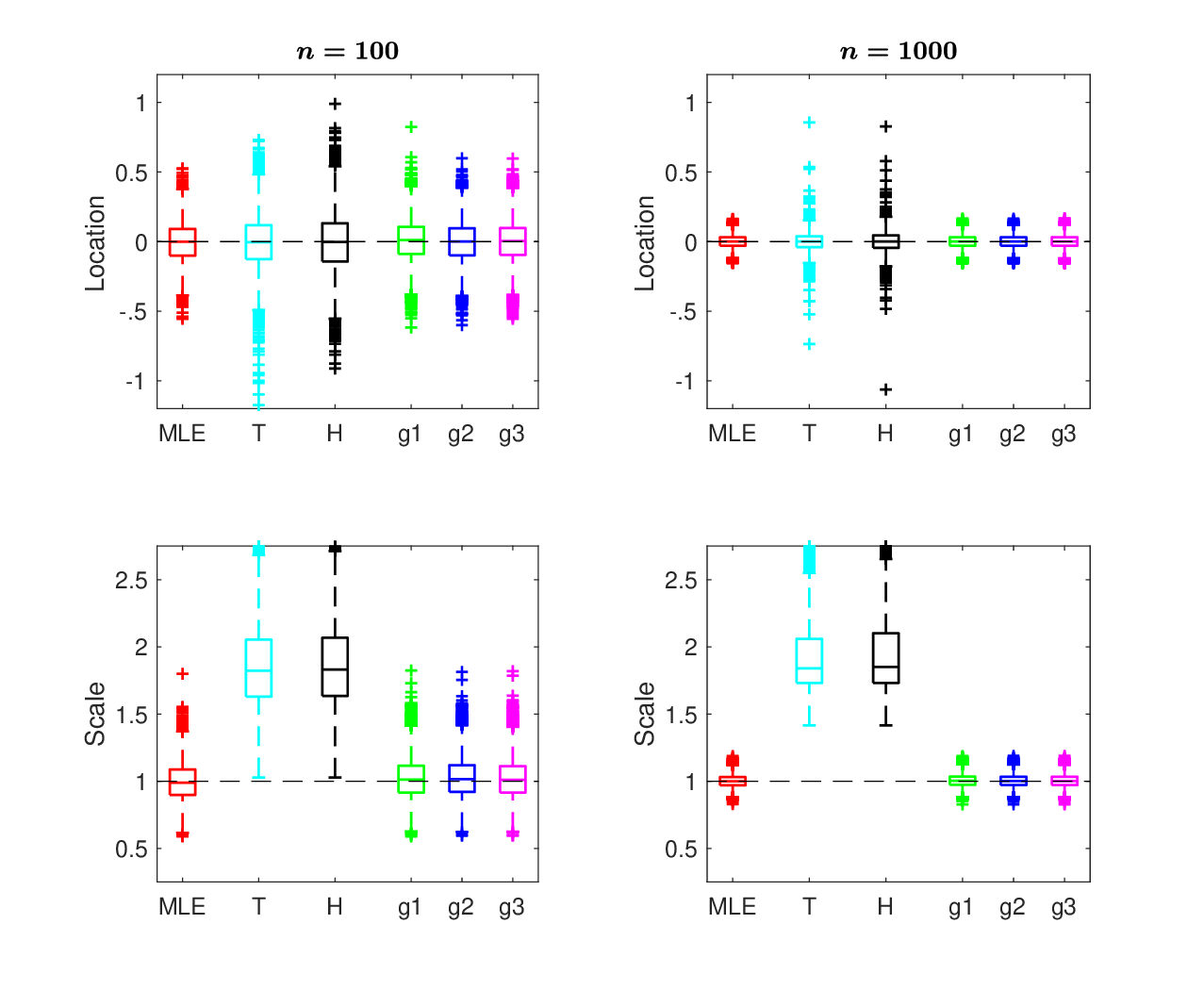}}
\\[-2ex]
{\sc Figure 4.1.} ~Boxplots of $\widehat{\mu}$ and $\widehat{\sigma}$ 
for Normal (top two rows) and Cauchy (bottom two rows) 
\\[-1ex]
distributions, using MLE, T, H, and gQLS (`g') estimators, 
where $(a, b)$ is equal to 
\\[-1ex]
$(0.02, 0.98)$ for g1, 
$(0.05, 0.95)$ for g2, 
and $(0.10, 0.90)$ for g3.
\end{center}

\begin{center}
\resizebox{160mm}{100mm}{\includegraphics{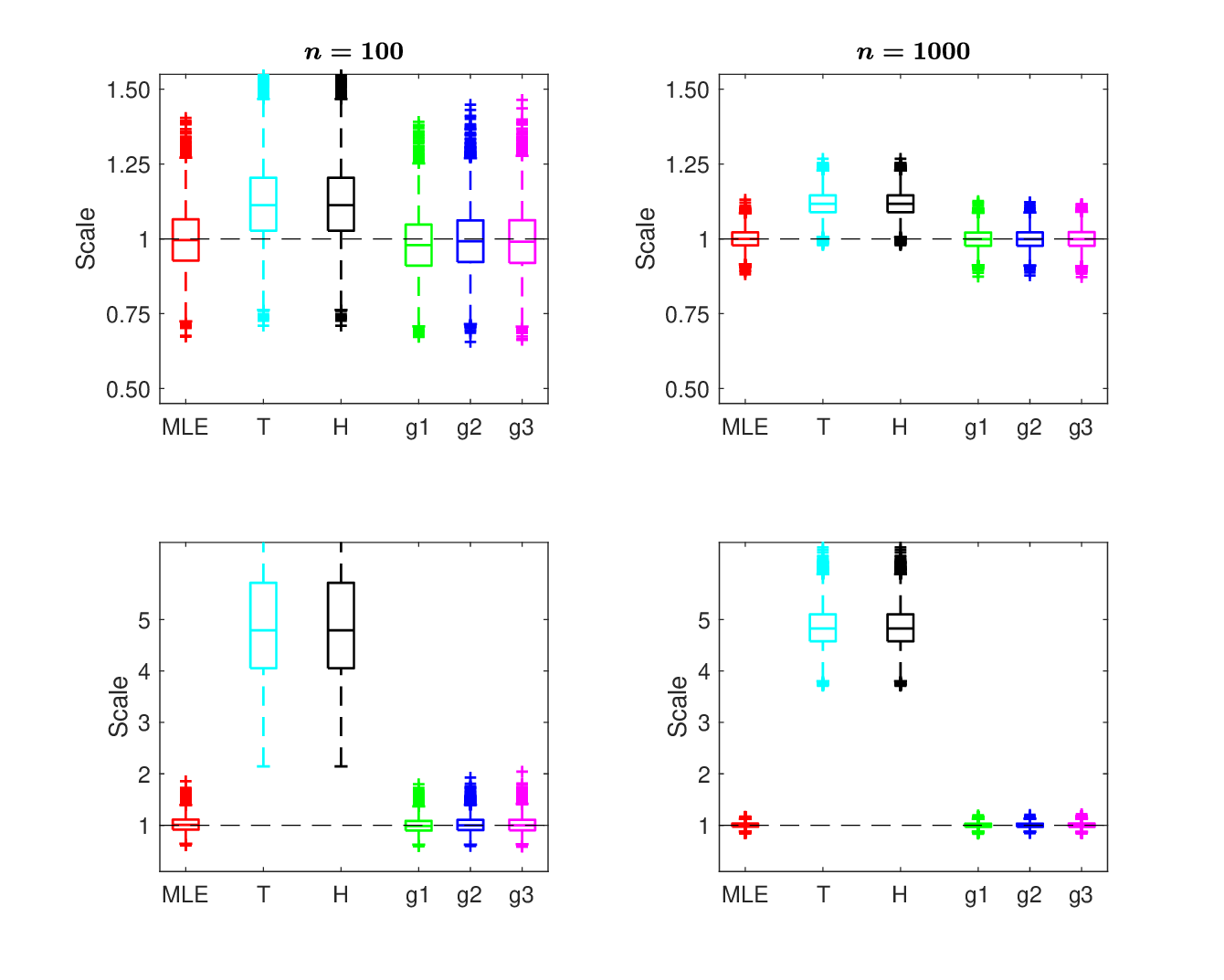}}
\\[-2ex]
{\sc Figure 4.2.} ~Boxplots of $\widehat{\sigma}$ for 
Exponential (top) and L{\'{e}}vy (bottom) distributions, 
\\[-1ex]
using MLE, T, H, and gQLS (`g') estimators, 
where $(a, b)$ is equal to 
\\[-1ex]
$(0.02, 0.98)$ for g1, 
$(0.05, 0.95)$ for g2, 
and $(0.10, 0.90)$ for g3.
\end{center}

\medskip

It is also of interest to see how fast these estimators can be 
computed when sample sizes are very large, which is quite common 
nowadays. Using $\mbox{MATLAB}^{\copyright}$ R2024b software on 
a basic laptop (with Apple M2 8-core CPU, RAM 8GB, Mac OS), the MLE, 
T, H and gQLS estimators of six location-scale families have been 
computed for samples of size $n = 10^6, 10^7, 10^8, 10^9$ and their 
computational times (in seconds) have been recorded in Table 4.1. 
Note that for all these distributions, the gQLS estimators have 
explicit formulas (although they require inversion of medium-sized 
matrices), MLE has explicit formulas for four chosen distributions 
but requires numerical optimization for Cauchy and Logistic, and 
the T and H estimators are always computed iteratively.
The conclusion is clear: the computational costs for gQLS are on a par 
with the explicit-formula MLEs and at least 10 times less than those 
of optimization-based MLEs. The T and H estimators perform similarly 
to non-explicit MLEs. The computational advantage of gQLS is highly 
relevant in situations involving ``big data''.

\newpage

\begin{center}
{\sc Table 4.1.} Computational times (in seconds) of 
MLE, T, H, and gQLS for large $n$.

\medskip

\begin{tabular}{|c|c|cccccc|}
\hline
Sample & Estimation & 
\multicolumn{6}{|c|}{Probability Distribution} \\[-0.5ex]
Size & Method & 
Cauchy & Exponential & Laplace & L{\'{e}}vy & Logistic & Normal \\
\hline
$n = 10^6$ & MLE & 2.22 & 0.01 & 0.06 & 0.05 & 0.67 & 0.02 \\
 & T & 0.75 & 0.73 & 0.59 & 0.65 & 0.70 & 0.60 \\
 & H & 0.53 & 0.46 & 0.54 & 0.50 & 0.49 & 0.41 \\
 & gQLS & {\bf 0.04} & {\bf 0.08} & {\bf 0.06} & {\bf 0.07} & 
{\bf 0.04} & {\bf 0.06} \\
\hline
$n = 10^7$ & MLE & 19.15 & 0.07 & 0.19 & 0.29 & 4.57 & 0.20 \\
 & T & 6.18 & 6.12 & 5.13 & 3.02 & 3.56 & 4.04 \\
 & H & 4.95 & 4.22 & 3.77 & 2.80 & 3.85 & 3.06 \\
 & gQLS & {\bf 0.28} & {\bf 0.32} & {\bf 0.33} & {\bf 0.46} & 
{\bf 0.25} & {\bf 0.38} \\
\hline
$n = 10^8$ & MLE & 414.16 & 0.96 & 1.78 & 3.33 & 76.61 & 2.44 \\
 & T & 180.00 & 158.60 & 142.02 & 67.95 & 133.52 & 123.77 \\
 & H & 157.10 & 106.63 &119.39 & 60.75 & 106.09 & 105.90 \\
 & gQLS & {\bf 3.10} & {\bf 2.97} & {\bf 3.59} & {\bf 5.11} & 
{\bf 2.52} & {\bf 4.18} \\
 \hline
$n = 10^9$ & MLE & $**$ & 154 & 844 & 531 & 28461 & 625 \\
 & T & ** & ** & ** & ** & ** & ** \\
 & H & ** & ** & ** & ** & ** & ** \\
 & gQLS & {\bf 434} & {\bf 484} & {\bf 702} & {\bf 751} & 
{\bf 469} & {\bf 798} \\
\hline
\multicolumn{8}{l}{\footnotesize $**$ ~For $n=10^9$, T and H and 
MLE at Cauchy failed to converge over a several days span.} \\
\end{tabular}
\end{center}

\subsection{Good Data, Bad Data}

When the distributional assumption is correct (``clean'' or ``good'' 
data scenario), the simulated large-sample performance of MLE or 
gQLS is consistent with the asymptotic results of Section 3.2, which 
was verified in the previous section. The H and T estimators perform 
well under the Normal case but require retuning for Cauchy and other 
distributions. When data are contaminated by outliers (``bad'' data 
scenario), all estimators are affected by it, but to a different 
extent. As is evident from the boxplots of Figure 4.3, the robust 
gQLS estimators successfully cope with outliers as long as their 
breakdown point exceeds the level of contamination $\varepsilon$.
And H and T perform very well under Normal, as expected. 
Robust estimators work especially well for estimation of $\mu$ and 
less so for estimation of $\sigma$. Further, for estimation of 
$\sigma$ under Normal, MLEs completely miss the target and their 
variability gets significantly inflated even for the smallest 
levels of contamination. For Cauchy, which easily accommodates 
the outliers from $\mbox{Normal} \, (\mu^* = 1, \sigma^* = 3)$, MLEs 
perform reasonably well. This suggests that to mitigate the effects 
of potential contamination on MLEs, a prudent approach is to always 
assume that data follow a heavy-tailed distribution. Of course, if 
one were to take that route they would have to accept the fact that 
no mean and other moments exist, and thus all subsequent inference 
should be based on quantiles.
Finally, more simulations have been conducted using Laplace, Logistic, 
and other distributions. The conclusions were similar to those of 
Normal and Cauchy: if a light-tailed distribution is assumed, 
contamination is devastating to MLEs, but a heavier-tailed 
distribution can correct the MLEs' performance. 
Those additional studies will not be presented here.

\subsection{Goodness of Fit}

A simulation study has been conducted to assess the power properties 
of the goodness-of-fit tests based on statistics $W$ and 
$W_{\mbox{\tiny out}}$. The results are presented in Tables 4.2-4.3 
(for $W$) and 4.4 (for $W_{\mbox{\tiny out}}$). 
The following conclusions emerge from the tables.
\begin{itemize}
  \item For the test based on $W$ (Table 4.2), the estimated 
probability of rejecting the null hypothesis approaches 1 as 
$n \rightarrow 1000$ for most distributions under $H_0$ and 
most alternatives. The exceptions are Logistic against Normal 
and $F_{0.05}$, and Normal against Logistic and $F_{0.05}$. 
For $n=100$, Cauchy has very low power against all alternatives,
and the test designed for Cauchy exceeds its level of 0.05. 
To make the test \eqref{gof1} maintain the intended 
significance level, we recommend using simulated thresholds 
instead of chi-square quantiles. In Table 4.3, the results of 
a simulation study based on $W$, with simulated thresholds, 
show that this modification indeed corrects the problem.
Comparisons between different levels of $(a, b)$, i.e., 
$(0.02, 0.98)$ versus $(0.05, 0.95)$ versus $(0.10, 0.90)$, 
do reveal some patterns. However, recall that choosing one 
pair of $(a,b)$ versus another means comparing the model fit 
on two overlapping but different ranges of data.

  \item For the test based on $W_{\mbox{\tiny out}}$ (Table 4.3),
all model fits are compared on the same set of quantiles, 
ranging from the level 0.01 to 0.99 (50 quantiles in total). 
The estimated probability of rejecting the null hypothesis 
approaches 1 as $n \rightarrow 1000$ for most distributions 
under $H_0$ and most alternatives. The power of Logistic 
against Normal and $F_{0.05}$ is still low, but higher than 
that based on $W$. This time Normal exhibits fairly high power 
against Logistic and $F_{0.05}$. The patterns among different 
choices of $(a, b)$ are mixed and depend on $H_0$ and the 
alternative distribution. All tests match the significance
level of 0.05. Interestingly, for $n=100$ the Cauchy-based 
test has no power at all against any of the selected 
alternatives.
\end{itemize}

\newpage

\begin{center}
\resizebox{160mm}{100mm}{\includegraphics{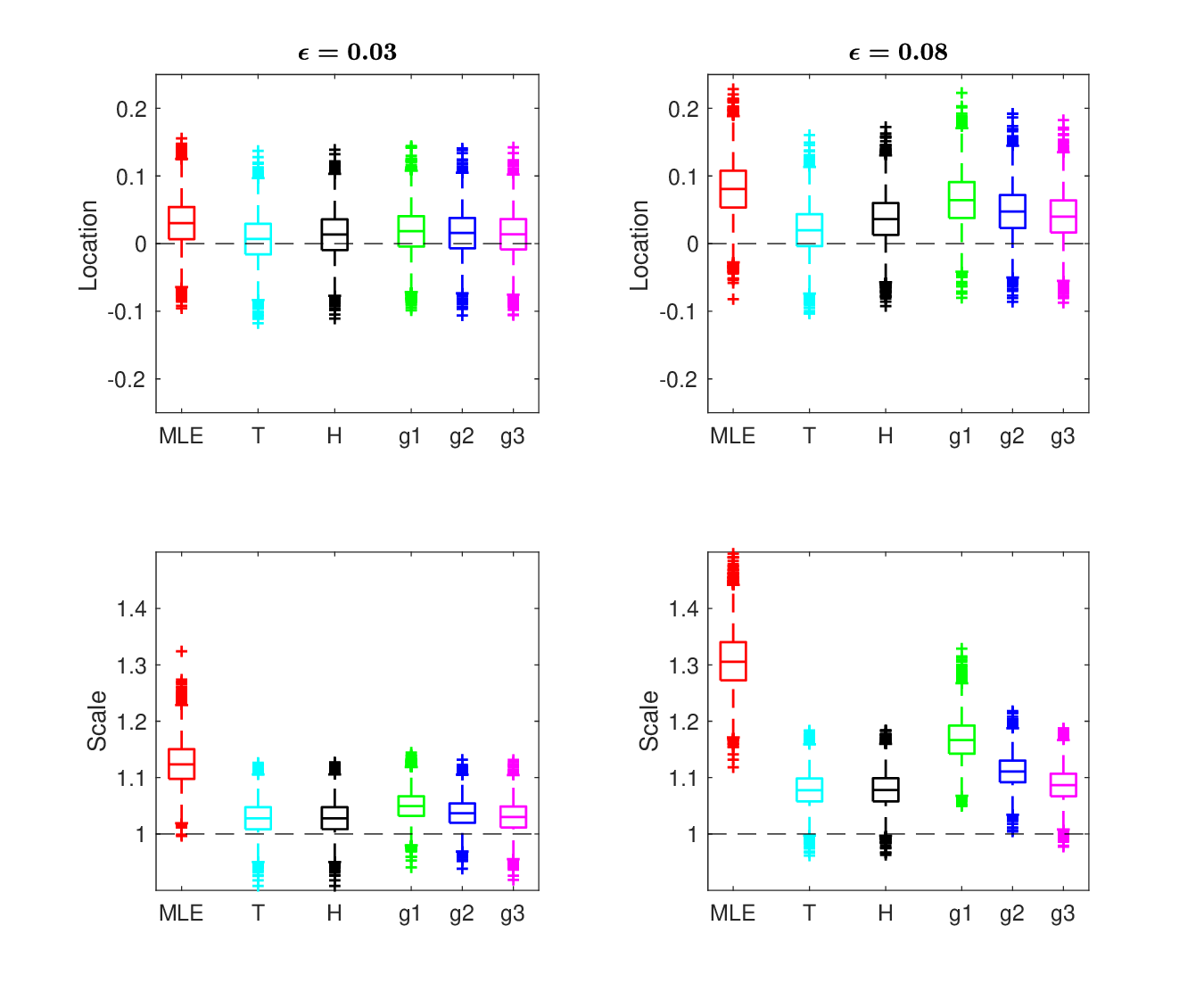}}
\resizebox{160mm}{100mm}{\includegraphics{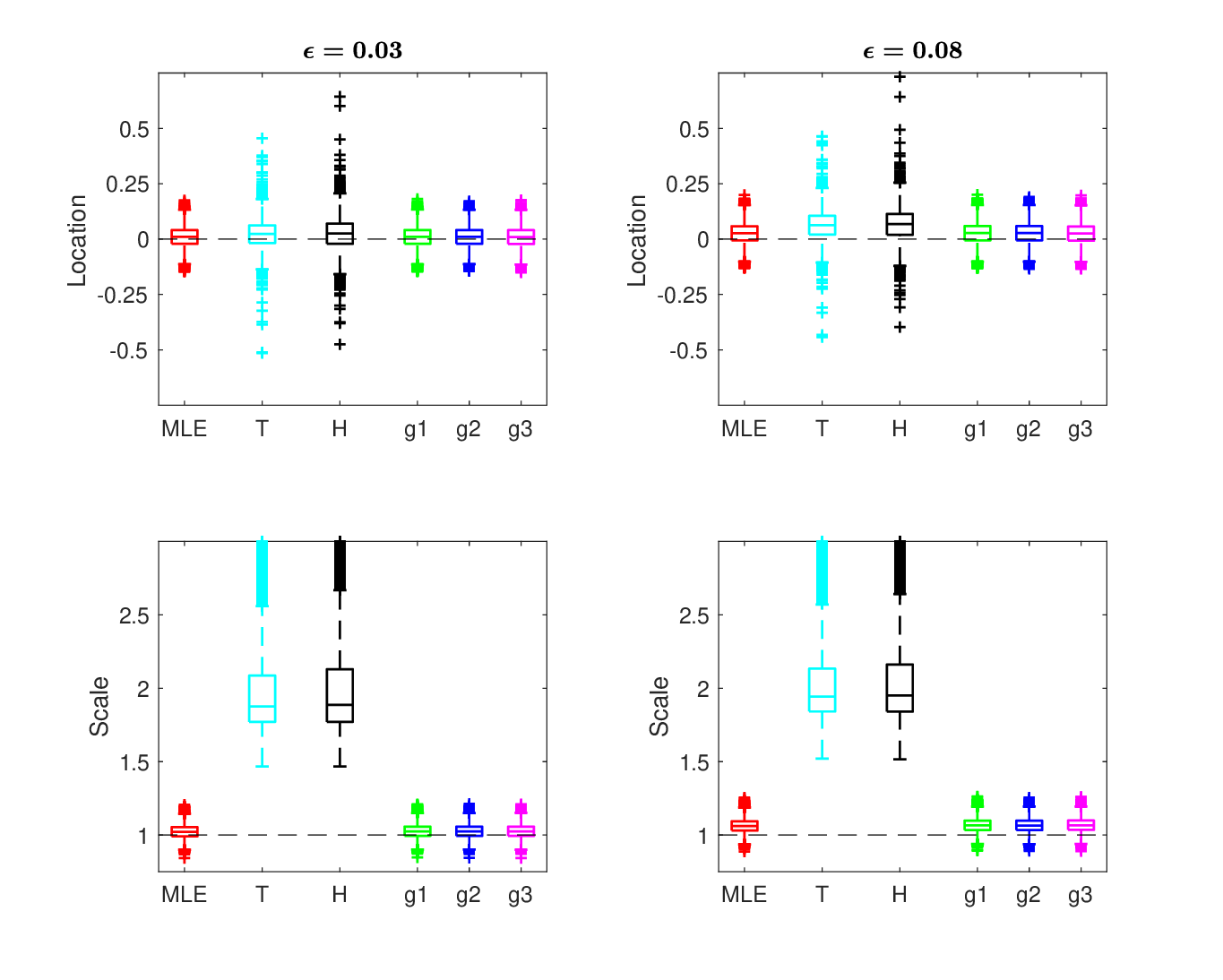}}
\\[-2ex]
{\sc Figure 4.3.} ~Boxplots of $\widehat{\mu}$ and 
$\widehat{\sigma}$ for Normal (top two rows) and Cauchy 
(bottom two rows) 
\\[-1ex]
distributions, using MLE, T, H, and gQLS (`g') estimators, 
where $(a, b)$ is equal to 
\\[-1ex]
$(0.02, 0.98)$ for g1, 
$(0.05, 0.95)$ for g2, 
and $(0.10, 0.90)$ for g3.
\end{center}

\newpage

\begin{center}
{\sc Table 4.2.} Proportions of rejections of $H_0$ by the 
goodness-of-fit test \eqref{gof1} at $\alpha = 0.05$ for
\\[-1ex]
several distributions under $H_0$ and $H_A$, and varying $n$. 
In all cases, $\mu = 0$ and $\sigma = 1$,
\\[-1ex]
and $F_{0.05} = 0.95 \, \mbox{Normal} \, (\mu = 0, \sigma = 1) + 
0.05 \, \mbox{Normal} \, (\mu^* = 1, \sigma^* = 3)$.

\medskip

\begin{tabular}{|c|c|cccccc|}
\hline
gQLS & Assumed & \multicolumn{6}{|c|}{Data Generated by} \\[-0.5ex]
Estimator & Distribution ($H_0$) &
Cauchy & Gumbel & Laplace & Logistic & Normal & $F_{0.05}$ \\
\hline
\multicolumn{8}{l}{Sample Size: $n=100$} \\
\hline
$a=0.02, b=0.98$ & Cauchy & {\bf 0.25} & 0.08 & 0.04 & 0.05 & 0.07 & 0.06 \\
 & Gumbel   & 1.00 & {\bf 0.08} & 0.89 & 0.66&0.47 & 0.51\\
 & Laplace  & 0.96 & 0.43 & {\bf 0.09}  & 0.21 & 0.34 & 0.27\\
 & Logistic & 1.00 & 0.32 & 0.28 & {\bf 0.09}  & 0.09 & 0.13 \\
 & Normal   & 1.00 & 0.46 & 0.52 & 0.16 & {\bf 0.08}  & 0.20 \\
\hline
$a=0.05, b=0.95$ & Cauchy & {\bf 0.18}  & 0.08 & 0.04 & 0.05 & 0.07 & 0.06 \\
 & Gumbel & 0.99 & {\bf 0.07}  & 0.75 & 0.45 & 0.32 & 0.32 \\
 & Laplace & 0.78 & 0.37 & {\bf 0.08}  & 0.17 & 0.26 & 0.21\\
 & Logistic & 0.96 & 0.29 & 0.23 & {\bf 0.08}  & 0.08 & 0.07 \\
 & Normal & 0.98 & 0.37 & 0.38 & 0.11 & {\bf 0.08}  & 0.09 \\

\hline
$a=0.10, b=0.90$ & Cauchy & {\bf 0.14}  & 0.11 & 0.06 & 0.08 & 0.09 & 0.08 \\
 & Gumbel & 0.89 & {\bf 0.08}  & 0.56 & 0.30 & 0.24 & 0.23 \\
 & Laplace & 0.40 & 0.28 & {\bf 0.09}  & 0.16 & 0.22 & 0.18 \\
 & Logistic & 0.74 & 0.22 & 0.20 & {\bf 0.09}  & 0.09 & 0.08 \\
 & Normal & 0.82 & 0.25 & 0.28 &0.10 & {\bf 0.08}  & 0.09 \\
\hline
\multicolumn{8}{l}{Sample Size: $n=1000$} \\
\hline
$a=0.02, b=0.98$ & Cauchy & {\bf 0.08}  & 1 &0.99 & 1 & 1 & 1 \\
 & Gumbel & 1 & {\bf 0.06}  & 1 & 1 & 1 & 1 \\
 & Laplace & 1 & 1 & {\bf 0.06}  & 0.97 & 1 & 0.99 \\
 & Logistic & 1 & 1 & 0.98 & {\bf 0.05}  & 0.35 & 0.18 \\
 & Normal & 1 & 1 & 1 & 0.57 & {\bf 0.05}  & 0.53 \\
\hline
$a=0.05, b=0.95$ & Cauchy & {\bf 0.07}  & 1 & 0.99 & 1 & 1 & 1 \\
 & Gumbel & 1 & {\bf 0.05}  & 1 & 1 & 1 & 1 \\
 & Laplace & 1 & 1 & {\bf 0.05} & 0.94 & 1 & 1 \\
 & Logistic & 1 & 1 & 0.96 & {\bf 0.05} & 0.19 & 0.10 \\
 & Normal & 1 & 1 & 1 & 0.29 & {\bf 0.05} & 0.13 \\
\hline
$a=0.10, b=0.90$ & Cauchy & {\bf 0.06} & 1 & 0.67 & 1 & 1 & 1 \\
 & Gumbel & 1 & {\bf 0.05} & 1 & 1 & 1 & 0.99 \\
 & Laplace & 0.99 & 1 & {\bf 0.05} & 0.81 & 0.98 & 0.95 \\
 & Logistic & 1 & 1 & 0.86 & {\bf 0.05} & 0.10 & 0.07 \\
 & Normal & 1 & 1 & 0.99 & 0.13 & {\bf 0.05} & 0.07 \\
\hline
\end{tabular}
\end{center}

\newpage

\begin{center}
{\sc Table 4.3.} Proportions of rejections of $H_0$ by the 
goodness-of-fit test \eqref{gof1}, {\em with simulated thresholds\/}, 
\\[-1ex]
at $\alpha = 0.05$ for several distributions under $H_0$ and $H_A$, 
and varying $n$. In all cases, $\mu = 0$ and $\sigma = 1$,
\\[-1ex]
and $F_{0.05} = 0.95 \, \mbox{Normal} \, (\mu = 0, \sigma = 1) + 
0.05 \, \mbox{Normal} \, (\mu^* = 1, \sigma^* = 3)$.

\medskip

\begin{tabular}{|c|c|cccccc|}
\hline
gQLS & Assumed & \multicolumn{6}{|c|}{Data Generated by} \\[-0.5ex]
Estimator & Distribution ($H_0$) &
Cauchy & Gumbel & Laplace & Logistic & Normal & $F_{0.05}$ \\
\hline
\multicolumn{8}{l}{Sample Size: $n=100$} \\
\hline
$a=0.02, b=0.98$ & Cauchy & {\bf 0.05} & 0.00 & 0.00 & 0.00 & 0.00 & 0.00 \\
 & Gumbel   & 1.00 & {\bf 0.05} & 0.86 & 0.60 & 0.38 & 0.44\\
 & Laplace  & 0.94 & 0.30 & {\bf 0.05}  & 0.13 & 0.23 & 0.18\\
 & Logistic & 0.99 & 0.21 & 0.19 & {\bf 0.05}  & 0.05 & 0.08 \\
 & Normal   & 1.00 & 0.36 & 0.43 & 0.12 & {\bf 0.05}  & 0.15 \\
\hline
$a=0.05, b=0.95$ & Cauchy & {\bf 0.05}  & 0.00 & 0.00 & 0.00 & 0.00 & 0.00 \\
 & Gumbel & 0.99 & {\bf 0.05}  & 0.71 & 0.39 & 0.25 & 0.25 \\
 & Laplace & 0.74 & 0.26 & {\bf 0.05}  & 0.11 & 0.18 & 0.15\\
 & Logistic & 0.95 & 0.21 & 0.18 & {\bf 0.05}  & 0.05 & 0.05 \\
 & Normal & 0.98 & 0.30 & 0.32 & 0.08 & {\bf 0.05}  & 0.06 \\

\hline
$a=0.10, b=0.90$ & Cauchy & {\bf 0.05}  & 0.02 & 0.01 & 0.01 & 0.01 & 0.01 \\
 & Gumbel & 0.85 & {\bf 0.05}  & 0.47 & 0.23 & 0.17 & 0.16 \\
 & Laplace & 0.32 & 0.18 & {\bf 0.05}  & 0.09 & 0.13 & 0.12 \\
 & Logistic & 0.67 & 0.15 & 0.14 & {\bf 0.05}  & 0.05 & 0.05 \\
 & Normal & 0.77 & 0.17 & 0.20 &0.06 & {\bf 0.05}  & 0.05 \\
\hline
\multicolumn{8}{l}{Sample Size: $n=1000$} \\
\hline
$a=0.02, b=0.98$ & Cauchy & {\bf 0.05}  & 1 & 0.99 & 1 & 1 & 1 \\
 & Gumbel & 1 & {\bf 0.05}  & 1 & 1 & 1 & 1 \\
 & Laplace & 1 & 1 & {\bf 0.05}  & 0.97 & 1 & 1 \\
 & Logistic & 1 & 1 & 0.98 & {\bf 0.05}  & 0.33 & 0.16 \\
 & Normal & 1 & 1 & 1 & 0.57 & {\bf 0.05}  & 0.52 \\
\hline
$a=0.05, b=0.95$ & Cauchy & {\bf 0.05}  & 1 & 0.97 & 1 & 1 & 1 \\
 & Gumbel & 1 & {\bf 0.05}  & 1 & 1 & 1 & 1 \\
 & Laplace & 1 & 1 & {\bf 0.05} & 0.93 & 1 & 1 \\
 & Logistic & 1 & 1 & 0.95 & {\bf 0.05} & 0.19 & 0.09 \\
 & Normal & 1 & 1 & 1 & 0.29 & {\bf 0.05} & 0.12 \\
\hline
$a=0.10, b=0.90$ & Cauchy & {\bf 0.05} & 1 & 0.62 & 1 & 1 & 1 \\
 & Gumbel & 1 & {\bf 0.05} & 1 & 1 & 1 & 0.99 \\
 & Laplace & 0.99 & 1 & {\bf 0.05} & 0.81 & 0.98 & 0.95 \\
 & Logistic & 1 & 1 & 0.85 & {\bf 0.05} & 0.10 & 0.07 \\
 & Normal & 1 & 1 & 0.99 & 0.13 & {\bf 0.05} & 0.06 \\
\hline
\end{tabular}  
\end{center}

\newpage

\begin{center}
{\sc Table 4.4.} Proportions of rejections of $H_0$ by the 
goodness-of-fit test \eqref{gof2} at $\alpha = 0.05$ for
\\[-1ex]
several distributions under $H_0$ and $H_A$, and varying $n$. 
In all cases, $\mu = 0$ and $\sigma = 1$,
\\[-1ex]
and $F_{0.05} = 0.95 \, \mbox{Normal} \, (\mu = 0, \sigma = 1) + 
0.05 \, \mbox{Normal} \, (\mu^* = 1, \sigma^* = 3)$.

\medskip

\begin{tabular}{|c|c|cccccc|}
\hline
gQLS & Assumed & \multicolumn{6}{|c|}{Data Generated by} \\[-0.5ex]
Estimator & Distribution ($H_0$) &
Cauchy & Gumbel & Laplace & Logistic & Normal & $F_{0.05}$ \\
\hline
\multicolumn{8}{l}{Sample Size: $n=100$} \\
\hline
$a=0.02, b=0.98$ & Cauchy & {\bf 0.05} & 0 & 0 & 0 & 0 & 0 \\
 & Gumbel   & 1 & {\bf 0.05}  & 0.83 & 0.57 & 0.32 & 0.52 \\
 & Laplace  & 0.96 & 0.18 & {\bf 0.05}  & 0.08 & 0.14 & 0.15 \\
 & Logistic & 1 & 0.12 & 0.19 & {\bf 0.05}  & 0.04 & 0.14 \\
 & Normal   & 1 & 0.25 & 0.43 & 0.14 & {\bf 0.05}  & 0.32 \\
\hline
$a=0.05, b=0.95$ & Cauchy & {\bf 0.05}  & 0 & 0 & 0 & 0 & 0 \\
 & Gumbel & 1 & {\bf 0.05}  & 0.90 & 0.68 & 0.41 & 0.60 \\
 & Laplace & 0.98 & 0.13 & {\bf 0.05} & 0.06 & 0.10 & 0.13 \\
 & Logistic & 1 & 0.10 & 0.21 & {\bf 0.05}  & 0.03 & 0.15\\
 & Normal & 1 & 0.26 & 0.53 & 0.19 & {\bf 0.05}  & 0.37 \\
\hline
$a=0.10, b=0.90$ & Cauchy & {\bf 0.05}  & 0 & 0 & 0 & 0 & 0 \\
 & Gumbel & 1 & {\bf 0.05}  & 0.94 & 0.75 & 0.46 & 0.65 \\
 & Laplace & 0.99 & 0.07 & {\bf 0.06}  & 0.03 & 0.04 & 0.07 \\
 & Logistic & 1 & 0.09 & 0.28 & {\bf 0.06}  & 0.02  & 0.13 \\
 & Normal & 1& 0.28 & 0.66 & 0.24 & {\bf 0.05}  & 0.38  \\
\hline
\multicolumn{8}{l}{Sample Size: $n=1000$} \\
\hline
$a=0.02, b=0.98$ & Cauchy & {\bf 0.05}  & 1 & 0.48 & 1 & 1 & 1\\
 & Gumbel & 1 & {\bf 0.05}  & 1 & 1 & 1 & 1 \\
 & Laplace & 1 & 1 & {\bf 0.05}  & 0.89 & 1 & 0.99 \\
 & Logistic & 1 & 1 & 0.95 & {\bf 0.06}  & 0.24 & 0.45 \\
 & Normal & 1 & 1 & 1 & 0.59 & {\bf 0.05}  & 0.87 \\
\hline
$a=0.05, b=0.95$ & Cauchy & {\bf 0.05}  & 1 & 0.47 & 1 & 1 & 1 \\
 & Gumbel & 1 & {\bf 0.05}  & 1 & 1 & 1 & 1 \\
 & Laplace & 1 & 1 & {\bf 0.05}  & 0.87 & 1 & 0.99 \\
 & Logistic & 1 & 1 & 0.96 & {\bf 0.05}  & 0.21 & 0.46 \\
 & Normal & 1 & 1 & 1 & 0.67 & {\bf 0.05}  & 0.90 \\
\hline
$a=0.10, b=0.90$ & Cauchy & {\bf 0.05}  & 1 & 0.44 & 0.99 & 1 & 1 \\
 & Gumbel & 1 & {\bf 0.05}  & 1 & 1 & 1 & 1 \\
 & Laplace & 1 & 1 & {\bf 0.05}  & 0.82 & 1 & 0.98 \\
 & Logistic & 1 & 1 & 0.98 & {\bf 0.05} & 0.18 & 0.43\\
 & Normal & 1 & 1 & 1 & 0.76 & {\bf 0.05} & 0.91\\
\hline
\end{tabular}
\end{center}

\section{Real Data Examples}

To illustrate how our proposed estimators and goodness-of-fit 
tests work on real data, we will use the daily stock returns of 
Alphabet Inc., the parent company of {\em Google\/}, for the 
period from January 2, 2020, to December 29, 2023. The stock
prices are available at the {\em Yahoo!\/}{\em Finance\/} website
{\tt https://finance.yahoo.com/quote/GOOG/}. The daily returns 
are calculated by subtracting the {\em opening\/} price from 
the {\em closing\/} price. Below are summary statistics for 
these data.

\begin{center}
\begin{tabular}{c|ccccc|cc}
Sample Size & 
\multicolumn{5}{|c|}{Five-Number Summary} &
\multicolumn{2}{|c}{Moments} \\[-0.5ex]
$n$ & $min$ & $q_1$ & $q_2$ & $q_3$ & $max$ & $mean$ & $std. dev.$ \\
\hline
$1006$ & -6.5275 & -0.8900 & $0.1487$ & $1.1095$ & $7.6735$ & 
$0.0905$ & $1.6661$ \\
\end{tabular}
\end{center}

\medskip

Since the histogram of these data is approximately symmetric,
a bell-shaped distribution should offer a good fit. Therefore, all the 
distributions of Section 4.4 were fitted to the daily returns using 
gQLS. Their probability-probability and data-probability plots (with 
$a = 0.02$ and $b =0.98$; the other choices of $a$ and $b$ result in 
virtually identical plots) are presented in Figure 5.1. As is evident 
from the figure, the Cauchy and Gumbel models are not appropriate for 
this dataset but Laplace, Logistic and Normal offer reasonably close 
fits. 

Further, according to the Tukey's outlier test, the daily returns 
that fall outside the interval
\[
\big[ q_1 - 1.5 \cdot (q_3-q_1); \, q_3 + 1.5 \cdot (q_3-q_1) \big] 
~=~ \big[ \mbox{-3.89}; \, 4.11 \big]
\]
should be labeled as potential outliers. Hence, this dataset has 
14 lower outliers 
(-6.53, -5.44, -5.39, -5.09, -5.08, -4.77, -4.64, -4.62, -4.51, 
-4.41, -4.39, -4.31, -4.24, -4.05) 
and
9 upper outliers 
(4.34, 4.41, 4.50, 4.57, 4.94, 6.33, 6.37, 6.53, 7.67), 
which corresponds to roughly 1\% contamination in each tail. Since 
we selected $a = 1-b \geq 0.02$, all the gQLS estimators used here 
offer sufficient protection against the non-parametrically identified 
outliers.

\begin{center}
\resizebox{170mm}{155mm}{\includegraphics{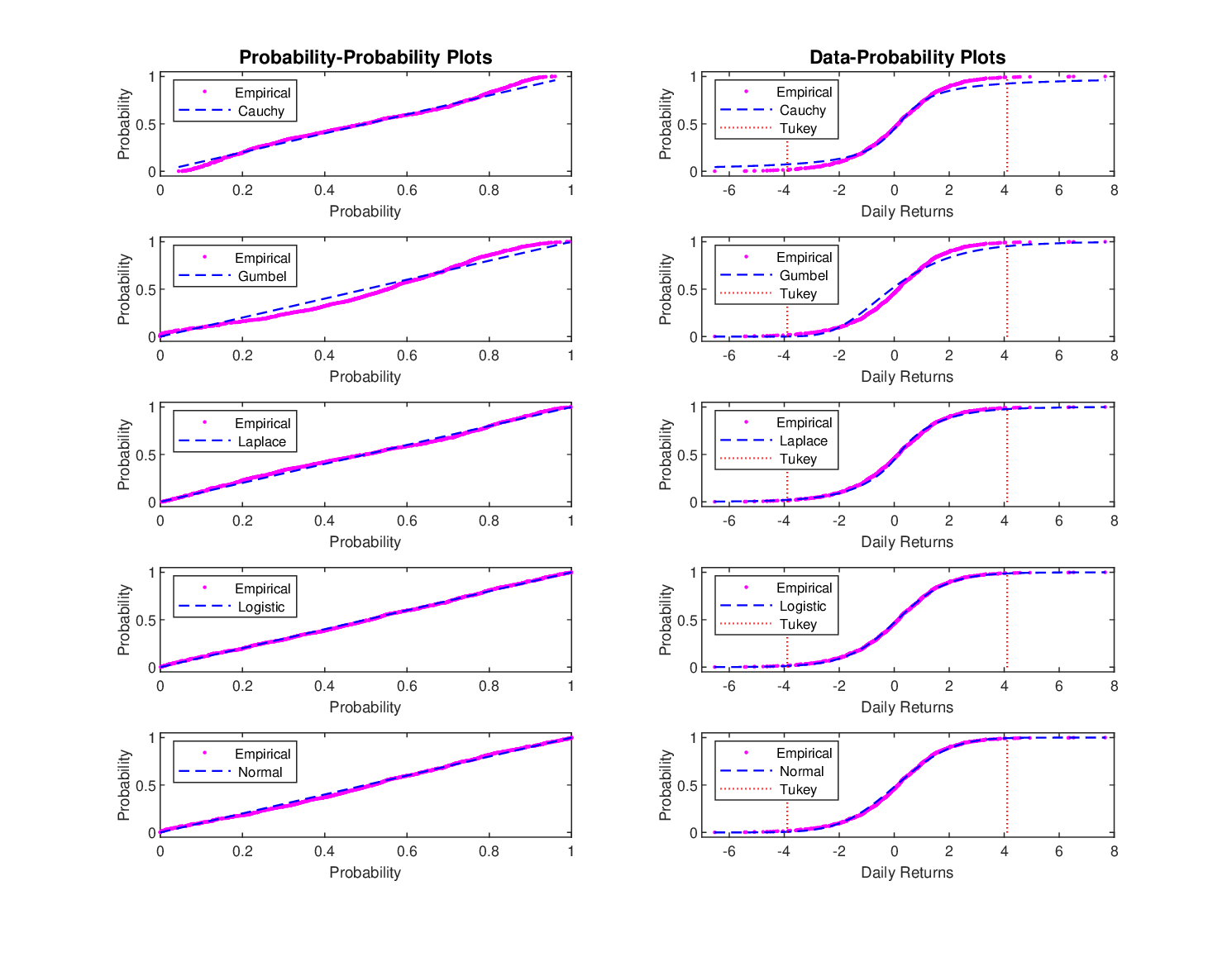}}
\\[-4ex]
{\sc Figure 5.1.} ~Probability plots for five location-scale 
families fitted to the daily returns
\\[-1ex]
of the Google stock (1/02/2020 -- 12/29/2023). 
According to the Tukey's test, 
\\[-1ex]
the daily returns that fall outside $[-3.89; \, 4.11]$ are 
labeled as outliers.
\end{center}

Furthermore, this preliminary analysis can be formally validated by
applying the goodness-of-fit tests \eqref{gof1}, {\em with simulated 
thresholds\/}, and \eqref{gof2}. The results are summarized in 
Table 5.1.
As is evident from the table, the Logistic distribution provides 
the best fit among the candidate models, with its $p$-values 
significantly exceeding the 0.10 level under both tests. Note 
that the more robust estimators (i.e., those with higher $a = 1-b$) 
achieve better fits according to the chi-square test \eqref{gof1}. 
This pattern is particularly evident in the case of the Normal 
distribution and $a=0.10, \, b=0.90$, but it is not surprising
because as $a = 1-b$ increases the quantile range for which 
residuals are calculated shrinks making the fit better. On 
the other hand, the test \eqref{gof2} computes residuals on 
the universal set of 50 quantile levels (from 0.01 to 0.99) and 
practically shows no sensitivity to the choice of $a$ and $b$.

Finally, it is important to keep in mind that although for the 
dataset considered here, we see a clear ``winner'' among the five 
fitted models, it would be prudent to retain the other two models 
-- Laplace and Normal -- for sensitivity checks, contract pricing, 
or other risk management applications.

\begin{center}
{\sc Table 5.1.} Parameter estimates and goodness-of-fit statistics
for several location-scale 
\\[-1ex]
families fitted to the daily returns of the Google stock 
(1/02/2020 -- 12/29/2023).

\medskip

\begin{tabular}{|c|c|cc|cc|}
\hline
gQLS Estimator & Assumed & \multicolumn{2}{|c|}{Parameter Estimates} & 
\multicolumn{2}{|c|}{Goodness-of-Fit Statistics} \\[-0.5ex]
(with $k = 25$) & Distribution & 
~~~~ $\widehat{\mu}$ ~~ & ~~~~ $\widehat{\sigma}$ ~~ & 
$W$ {\footnotesize ($p$-value)} & 
$W_{\mbox{\tiny out}}$ {\footnotesize ($p$-value)} \\
\hline
$a=0.02, b=0.98$ & Cauchy & 0.16 & 0.95 & 
76.74 {\footnotesize (0.00)} & 89.41 {\footnotesize (0.00)} \\
 & Gumbel   & -0.68 & 1.58 & 
235.78 {\footnotesize (0.00)} & 343.12 {\footnotesize (0.00)} \\
 & Laplace  &  0.15 & 1.27 & 
57.30 {\footnotesize (0.00)} & 71.11 {\footnotesize (0.03)} \\
 & Logistic &  0.11 & 0.91 & 
{\bf 31.11} {\footnotesize ({\bf 0.15})} & 
{\bf 45.45} {\footnotesize ({\bf 0.58})} \\
 & Normal   &  0.08 & 1.59 & 
44.01 {\footnotesize (0.01)} & 73.85 {\footnotesize (0.02)} \\
\hline
$a=0.05, b=0.95$ & Cauchy & 0.15 & 0.95 & 
75.66 {\footnotesize (0.00)} & 88.83 {\footnotesize (0.00)} \\
 & Gumbel   & -0.64 & 1.51 & 
167.95 {\footnotesize (0.00)} & 390.15 {\footnotesize (0.00)} \\
 & Laplace  &  0.15 & 1.30 & 
48.38 {\footnotesize (0.00)} & 68.40 {\footnotesize (0.04)} \\
 & Logistic &  0.12 & 0.91 & 
{\bf 26.86} {\footnotesize ({\bf 0.29})} & 
{\bf 44.70} {\footnotesize ({\bf 0.61})} \\
 & Normal   &  0.09 & 1.58 & 
36.17 {\footnotesize (0.06)} & 76.47 {\footnotesize (0.01)} \\
\hline
$a=0.10, b=0.90$ & Cauchy & 0.16 & 0.95 & 
60.02 {\footnotesize (0.00)} & 88.42 {\footnotesize (0.00)} \\
 & Gumbel   & -0.57 & 1.42 & 
112.89 {\footnotesize (0.00)} & 481.63 {\footnotesize (0.00)} \\
 & Laplace  &  0.15 & 1.33 & 
37.10 {\footnotesize (0.03)} & 66.73 {\footnotesize (0.05)} \\
 & Logistic &  0.12 & 0.91 & 
{\bf 21.82} {\footnotesize ({\bf 0.54})} & 
{\bf 45.45} {\footnotesize ({\bf 0.57})} \\
 & Normal   &  0.10 & 1.53 & 
{\bf 26.53} {\footnotesize ({\bf 0.28})} & 
87.28 {\footnotesize (0.00)} \\
\hline
\end{tabular}
\end{center}

\bigskip

\section{Concluding Remarks}

\subsection{Summary}

In this paper, two types of {\em quantile least squares\/} 
estimators for location-scale families have been introduced: 
ordinary (denoted oQLS) and generalized (denoted gQLS). Both
approaches are robust. While the oQLS estimators are quite 
effective for more general probability distributions 
(e.g., $g$-and-$h$ distributions), the gQLS estimators can 
match the levels of robustness of oQLS yet offer much higher 
efficiency for estimation of location and/or scale parameters. 
These properties have been derived analytically (for large $n$) 
and verified using simulations (for small and medium-sized samples). 
In addition, two goodness-of-fit tests have been contructed and 
their power properties have been investigated via simulations. 
Also, it has been established that computational times of these 
estimators are similar to those of explicit-formula MLEs, and 
are more than 10 times lower when MLEs have to be found using 
numerical optimization. For example, gQLS can be computed for 
a sample of a {\em billion\/} observations in 7-15 minutes while 
non-explicit MLEs, T, and H take more than 5 hours or do not 
converge at all.

\subsection{Practical Recommendations}

When implementing the QLS-type, T, and H (and other robust) 
estimators in practice, it is worthwhile to keep in mind 
the following points.

\begin{enumerate}
  \item {\em gQLS vs oQLS\/}. The gQLS estimators are theoretically 
stronger than the corresponding oQLS. The latter, however, are 
robust, consistent, and (being special cases of gQLSs) can be 
useful for finding initial values of gQLSs when these have to 
be computed recursively.

\vspace{-1ex}

  \item {\em Robustness, Efficiency, Outliers\/}. Many robust 
estimators have built-in tuning constants. For T and H, the 
standard choices of $c = 4.685$ and $k = 1.345$ lead to 
$\mbox{ARE} = 0.95$ for $\mu$ at the normal distribution and high 
robustness with $\mbox{BP} \approx 0.28$ (for joint estimation of 
$\mu$ and $\sigma$). But this does not automatically translate in 
their correct design for Cauchy, L{\'{e}}vy  or other models. They 
need to be retuned. For gQLS, the tuning constants are $a$ and $b$, 
and transitions between different location-scale families are seamless. 
Further, there is no explicit formula on how to choose $a$ and $b$. 
That is why it is important to study the estimators' trade-offs 
between efficiency ($0 \leq \mbox{ARE} \leq 1$, where 1 is best) 
and robustness ($0 \leq \mbox{ BP} \leq 0.50$, where 0.50 is best). 
Based on that analysis, the user has to decide on how much efficiency 
they are willing to sacrifice and then pick the most robust estimator 
that meets the ARE benchmark. For insightful illustrations on how this 
can be done in practice, we recommend a paper by 
Riani {\em et al\/}. (2020).
Alternatively, one can apply some nonparametric outlier screening
rule, get a sense of the level of contamination in the given 
dataset, and then select $a$ and $b$ accordingly (see Section 5). 
Our general rule of thumb: $a = 1-b \leq 0.10$ will yield 
$\mbox{LBP} = \mbox{UBP} \leq 0.10$ and $\mbox{ARE} \geq 0.80$ 
for gQLS estimators of $(\mu, \sigma)$ under most location-scale 
families, which is satisfactory when working with large datasets.

\vspace{-1ex}

  \item {\em Goodness-of-fit Tests}. The goodness-of-fit test 
\eqref{gof2} is more effective than \eqref{gof1}. Also, if the sample 
size is large, say $n \geq 10^4$, then it makes sense to use a larger 
set of validation quantiles. For example, a set of 99 quantile levels
($0.01, 0.02, \ldots, 0.98, 0.99$) may be considered.

\vspace{-1ex}

  \item {\em Data Analysis\/}. In real data analysis, the models with 
lower $p$-values (but visually acceptable fits) should be retained for 
sensitivity checks in further applications.
\end{enumerate}

\subsection{Future Outlook}

The research presented in this paper can be extended in several 
directions. The most obvious one is to develop QLS estimators
for more general classes of distributions, including modern
goodness-of-fit tests \citep[][]{kgm23}. Another direction is 
to follow the literature on $L$-moments 
\citep[][]{h90} and trimmed $L$-moments (see \citet[][]{es03}; 
\citet[][]{h07}) and construct QLS-type statistics to summarize 
the shapes of parametric distributions. This line of research 
is more probabilistic in nature. Further, on the robust statistics 
side, direct comparisons with the MTM estimators of \citet[][]{bjz09}, 
MWM estimators (see \citet[][]{zbg18a}; \citet[][]{zbg18b}), or 
methods based on tail extensions with log-regularly varying 
distributions \citep[][]{d15} are also of interest.
Finally, potential generalizations of the QLS approach to more 
complex data settings such as regression are possible but the user 
will have to figure out what feature of the underlying distribution 
can be ``explained''. For example, for symmetric location-scale 
families, $\mu$ is the most obvious candidate to be replaced with 
a linear combination of explanatory variables as it represents 
the center (i.e., median, mode, and in many cases the mean) of 
such distributions. But for asymmetric families such as L{\'{e}}vy, 
$\mu$ is the left endpoint of the distribution support. 
Is it worthwhile explaining it?

\section*{Acknowledgments}

The authors are very appreciative of valuable suggestions, 
thought-provoking questions, and useful comments provided by 
two anonymous referees, which helped to substantially improve 
the paper. Also, much of this work was completed while the first 
author was a Ph.D. student in the Department of Mathematical 
Sciences at the University of Wisconsin-Milwaukee.

\section*{Competing Interests}

The authors have no financial or proprietary interests in any 
material discussed in this article.

\end{document}